\documentclass[10pt,journal]{IEEEtran}
\usepackage[utf8]{inputenc}
\usepackage[T1]{fontenc}
\usepackage{microtype} 
\usepackage{cite}
\usepackage{graphicx}
\usepackage[caption=false]{subfig}
\usepackage{amsfonts,amsmath,amsthm,amssymb}
\usepackage{algorithm} 
\usepackage{algpseudocode}

\usepackage{amsmath} 

\usepackage{tikz}
\usetikzlibrary{automata,positioning}
\usetikzlibrary{decorations.pathreplacing,decorations.markings,shapes.geometric}
\usetikzlibrary{shapes,arrows}
\usetikzlibrary{backgrounds,calc,positioning}

\theoremstyle{definition}

\usepackage{mathtools}
\usepackage{commath}
\theoremstyle{remark}

\theoremstyle{plain}

\newcommand{\bmath}[1]{\mbox{\boldmath$#1$}}

\usepackage{cite}
\usepackage{hyperref}
\hypersetup{
        colorlinks = true,
        citecolor=red,
}

\ifCLASSINFOpdf
\else
\fi
\begin{document}
%

\allowdisplaybreaks

\title{Fair and Efficient Distributed Edge Learning with Hybrid Multipath TCP}

%
%
%
%
 \author{Shiva~Raj~Pokhrel, \textit{senior member}, IEEE, Jinho~Choi, \textit{senior member}, IEEE, and Anwar~Walid, \textit{Fellow}, IEEE
 \IEEEcompsocitemizethanks{\IEEEcompsocthanksitem S.~R.~Pokhrel and J.~Choi are with the School of Information Technology, Deakin University, Geelong. A.~Walid is with Amazon, New York, USA.
Email: \{shiva.pokhrel, jinho.choi\}@deakin.edu.au,anwar.i.walid@gmail.com.
}
}
\maketitle
\begin{abstract}
The bottleneck of distributed edge learning (DEL) over wireless has shifted from computing to communication, primarily the \textit{aggregation-averaging} (Agg-Avg) process of DEL. The existing transmission control protocol (TCP)-based data networking schemes for DEL are application-agnostic and fail to deliver adjustments according to application layer requirements. As a result, they introduce massive excess time and undesired issues such as unfairness and stragglers. Other prior mitigation solutions have significant limitations as they balance data flow rate from workers across paths but often incur imbalanced backlogs when the paths exhibit variance, causing stragglers. To facilitate a more productive DEL, we develop a hybrid multipath TCP (MPTCP)
by combining model-based and deep reinforcement learning (DRL) based MPTCP  for DEL that strives to realize quicker iteration of DEL and better fairness (by ameliorating stragglers). Hybrid MPTCP essentially integrates two radical  TCP developments: i) successful existing model-based MPTCP control strategies and ii) advanced emerging DRL-based techniques, and introduce a novel hybrid MPTCP data transport for easing the communication of \textit{Agg-Avg} process. Extensive emulation results demonstrate that the proposed hybrid MPTCP can overcome excess time consumption and ameliorate the application layer unfairness of DEL effectively  without injecting additional inconstancy and stragglers.
\end{abstract}

\maketitle
\IEEEdisplaynontitleabstractindextext

%
\IEEEpeerreviewmaketitle

\section{Background} 
\label{sec:introduction}

\label{sec:introduction}
%



With the growing advances of machine learning models as well as the ensuing surge of data acquisition approaches, the level of complex computing involved in learning has increased substantially. Intensive training for learning on a single machine has become entirely inefficient~\cite{luo2018parameter}. As a solution to this problem, distributed learning (DL)~\cite{park2021communication}, due to its decentralized nature, 
is gaining much more attention in wireless edge computing industry verticals 
than centralized learning~\cite{9261995, pokhrel2020federated}. This is driven by the three main features of the so-called \textit{distributed edge learning} (DEL), viz. i) utilization of existing and cost-effective devices and infrastructure, ii) better privacy and robustness to failure, and iii) capability to harness multiple machines and aggregate resources/bandwidth. 

Distributed training and inference deduction, on the other hand, also necessitate communication between wireless devices and edge servers through wireless links~\cite{park2019distilling, Verbraeken2020}. As a result, wireless channels' inconsistency and lack of radio resources would significantly impact distributed learning performance. This paper aims at developing a framework for efficient transport protocol design and optimization of the DEL.
\begin{figure}[t]
    \centering
    \includegraphics[scale=0.415]{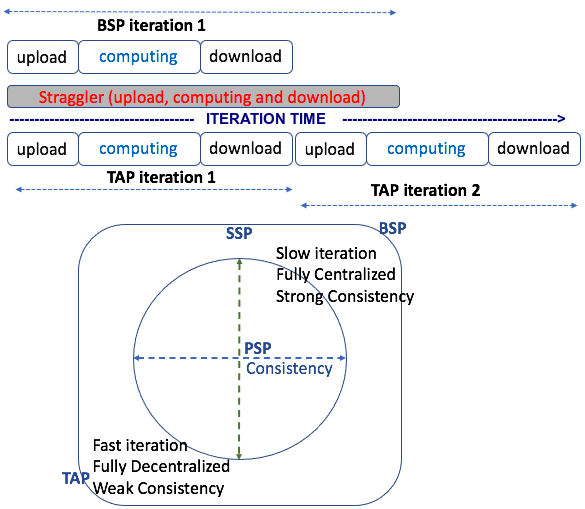}
    \caption{\color{black} {Different versions of data parallelism in distributed edge learning and their trade-offs over the iteration time. TAP works asynchronously in a completely decentralized fashion; however, its convergence and consistency are the lowest. SSP permits pioneering workers to move ahead of stragglers for
    a certain number of iterations. BSP strictly synchronizes in every iteration and guaranteed convergence.}}
    \label{fig:fig1}
\end{figure}

Data parallelism is the most widely used technique in DEL for its convenience~\cite{8756095, 9320519}. As shown in Fig.~\ref{fig:fig1}, a typical iteration of data parallelism in DEL involves three distinct stages (two communication and one computing): i) parameter/model downloading, ii) gradient/model computation, and iii) gradient/model uploading.

At the beginning of each iteration, each worker downloads the latest model or parameters from the global server and then computes to obtain
a new model and gradients using downloaded parameters and training data. Eventually, the server receives the gradients uploaded by the workers across the underlying network and (re)computes the model, which drives the workers to the next iteration. The data parallelism model explained above is the most generally practiced DEL method, usually referred to as a \textit{bulk synchronous parallelism model} (BSP)~\cite{chilimbi2014project}. 

{\color{black} In each iteration of the bulk synchronous model, all of the other workers involved in DEL need to wait for the slowest worker (not only due to computational delay but also due to communication delays and channel impairments (e.g., workers at low signal-to-noise ratio (SNR) area) or imbalanced bandwidth allocation among workers), referred to as \textit{straggler}~\cite{bitar2020stochastic, 9320519}), to accomplish its task.  To ameliorate such starvation effect due to straggler, various forms of parallelism model have been proposed in 
the literature, such as \textit{stale synchronous parallelism} (SSP)~\cite{yu2002design, zhao2019dynamic}, \textit{probabilistic synchronous parallelism} (PSP), \textit{total asynchronous parallelism} (TAP)~\cite{hsieh2017gaia, han2015giraph}, 
and so on (see~\cite{Verbraeken2020} and the references therein for a detailed comparison of parallelism schemes).} 

In Fig.~\ref{fig:fig1}, we compare different versions of data parallelism and summarize their trade-offs. In Fig.~\ref{fig:fig1}, the difference in data parallelism schemes relies on the different tolerances for their parameter
inconsistencies. As reported in~\cite{8756095, 9320519}, other prior mitigation solutions have significant limitations. They balance data flow rate from workers across paths but often incur backlogs (and bufferbloat) when the paths exhibit variance, causing stragglers. Moreover, to accelerate the training speed and maintain a high level of consistency, we have identified the following two bottleneck issues to be addressed:

\begin{enumerate}
    \item [$\circ$] \textit{Efficiency}. Minimize
the computing and communication (uploading and downloading latency) time of each worker. This is to speed up the overall iteration
process.
\item [$\circ$] \textit{Fairness}. Minimize
the gap between the straggler and pioneers. This is to attain better fairness and thereby speed up the overall distributed training and learning
process.
\end{enumerate}

To this end, our main focus is on developing
an effective solution to tackle the above two challenges from the data-networking perspective. 
This research direction is motivated by the recent observation of the continuous improvement in the computing power of 
devices, which has now completely shifted the bottleneck from computing to communication, in particular, the aggregation and averaging (Agg-Avg) process of 
DEL~\cite{tang2020communication, luo2018parameter, ouyang2020communication}. As in~\cite{pokhrel2018modeling, ouyang2020communication}, inefficient and unfair allocation of bandwidth and communication resources and/or channel impairments often leads to an undesirable delay in 
DEL (due to the straggler)~\cite{bitar2020stochastic, ouyang2020communication, 8756095, 9320519}.

In our earlier works~\cite{pokhrel2019fair, 7864463}, we have found that efficient and balanced bandwidth allocation across several TCP flows can greatly reduce the communication latency and ameliorate unfairness~\cite{xia2019rethinking}. Therefore, in this work, we investigate and design an intelligent and balanced bandwidth allocation scheme that can continuously learn to (potentially) boost the iterative training process and
the performance of DEL, with no (or minimal if required) 
modifications on the underlying network settings. We consider adopting the same (existing) parallelism schemes (recall Fig.~\ref{fig:fig1})  and focus on designing such an intelligent and load-balanced bandwidth allocation to address the following four critical aspects:

\begin{enumerate}
\item \textit{Transient relationship of data flows from multiple workers.} Existing data transport mechanisms design (such as TCP are independent by design) do not consider the temporal dependency between multiple workers/devices and their data flows. An intelligent bandwidth balancing algorithm should be able to
capture the temporal relationship of the data flows from the workers in an
online fashion to equalize the communication cost of all workers. This is
 to maximize long-term system utility and the performance of DEL (by minimizing the gap in communication cost between the straggler and pioneer workers).

\item \textit{Harnessing bandwidths from heterogeneous networks.} A new multipath data transport protocol such as Multipath TCP (MPTCP)~\cite{9444785, pokhrel2020multipath} has the potential to harness bandwidth from multiple links by splitting data flows into multiple flows. 
To take advantage of the increasingly available wireless access technologies and multipath capabilities, successful features of MPTCP can be inherited as an enabler for harnessing the  available bandwidths.
This is valuable to minimize the communication time and improve the performance of DEL.

\item \textit{Smoother and scalable bandwidth allocation.} Dynamic bandwidth allocation among multiple workers  is a delicate balancing act. As reported in~\cite{shafigh2006dynamic, wen2020joint}, shifting a large bandwidth from one flow to another is highly nontrivial at timescales relevant to the workers/devices. Moreover, the literature has recognized that the effects of dynamic switching of bandwidths among flows play an adverse role in the stability of the bandwidth allocation algorithms. {\color{black} This consequently affects convergence and consistency of the DEL. As a result, it is important to consider that the trajectory of the DEL process generated by a new data transport algorithm (such as MPTCP) should guarantee smooth convergence and stability towards global equilibrium. To this end, MPTCP scheduler governs how data packets are to be routed across multiple paths. }

\item \textit{Efficiency and Responsiveness.} It is known that a linearized system defines the responsiveness of multipath bandwidth allocation algorithm around equilibrium (how  fast  does  the  system  converge  to  the  equilibrium locally?)~\cite{peng2016multipath}.\footnote{At a higher conceptual level, the complexity of the responsiveness of the linearized system (how  fast  does  the  system  converge  to  the  equilibrium locally)  is  determined  by  the  real  parts  of  the  eigenvalues of its Jacobian.} An algorithm that makes prompt decisions without sacrificing its performance is said to be having higher efficiency.
\end{enumerate}
\begin{figure}[t]
\centering 
\includegraphics[width=2.7025 in]{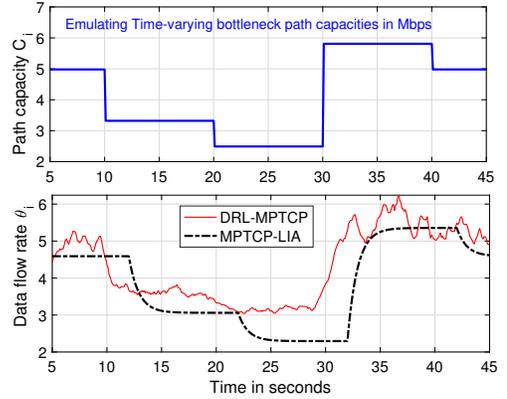}
\caption{\rm \textbf{Motivating observation 1} (at flow level): The top panel shows the variation in the bottleneck path capacity of one of the paths. The bottom panel illustrates the reaction of the workers using different variants of MPTCP to the variation in path capacity by changing their data flow rates. For the two paths, we fix the same delays ($50$ms), same path loss of $3\%$ and set bottleneck bandwidth of the other path ($8$~Mbps each) while continuously uploading or downloading files of $600$~MB each. We conduct several runs for \emph{DRL-based MPTCP and MPTCP-LIA} with the same setup by changing the path capacity $C_i$ of the bottleneck path. \label{fig:short}} 
\vspace{-5 mm}
\end{figure}
\section{Motivations and Contributions}

TCP has been the de-facto mechanism
for bandwidth allocation and congestion control  on the Internet for more than 30 years; however, in recent years, real momentum has been building up for
the major rethink on TCP. 
This is reflected, for example, by 
the work in recent years 
where Google researchers improve bottleneck bandwidth allocation (TCP BBR~\cite{cardwell2019bbrv2})\footnote{https://cloud.google.com/blog/products/networking/tcp-bbr-congestion-control-comes-to-gcp-your-internet-just-got-faster} to adapt to other new environments such as 5G networks using model-based approaches. 
{\color{black} In contrast to the classic model-based approach for TCP design~\cite{raiciu2011coupled, peng2016multipath, cardwell2019bbrv2},  DRL-based approaches~\cite{li2019smartcc,xu2019experience, pokhrel2020multipath, zhang2020machine}
can also be employed.} 

With DRL-based MPTCP~\cite{9444785}, such tuning and adaptation of bandwidth allocation protocol to new and time-varying environments can become an automatic process handled by the machine in few days (instead of years); thus avoiding the costly procedure of manually adapting or (re)engineering the bandwidth allocation and congestion control mechanism to specific scenarios. However, the application of
recent clean slate DRL-based design for DEL applications incurs its own challenges and costs. 
In this paper, we perform extensive studies to shed light on the problems and challenges of the clean slate DRL-based MPTCP designs.

\subsection{Motivating Observations}

\begin{figure}
    \centering
    \includegraphics[width=2.8025 in]{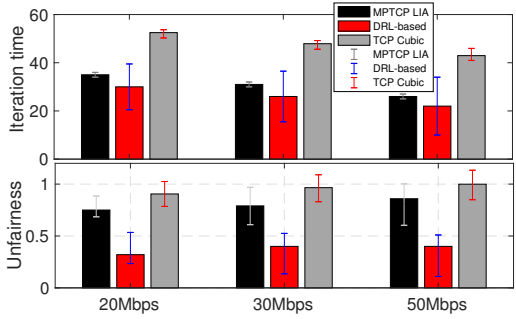}
    \caption{\rm{\color{black} \textbf{Motivating observation 2} (Marginal gain at the application level): With the setup and path capacity variation discussed in Fig.~\ref{fig:short}, we observe the average DEL iteration time and unfairness perceived by the workers using three different versions of TCP (DRL-based, MPTCP LIA,
and TCP CUBIC) for their data flows. Measurements were collected from three different experimental runs with the three different TCP versions separately.\label{fig:short2}}}
    \label{fig:my_label}
\end{figure}
It is known that TCP operates at flow level, which clearly implies an application-agnostic bandwidth
allocation~\cite{8756095, 9320519}. Along the lines of~\cite{8756095}, we consider TCP as a baseline for a motivating example (as it is widely used by DEL systems). We conduct an emulation experiment for comparing the performance of data flows from workers using MPTCP LIA~\cite{raiciu2011coupled}, DRL-based MPTCP~\cite{pokhrel2020multipath} over two paths (further
 details of the settings provided later in Sec.~\ref{sec:trainingandeval}). We fixed delays ($50$ms) and loss ($3\%$ path loss) and fixed capacity of one path ($8$~Mbps each) while uploading/downloading replicas of $600$~MB  files. We perform different runs for \emph{DRL-based MPTCP}~\cite{pokhrel2020multipath} and \emph{MPTCP-LIA}~\cite{raiciu2011coupled} with 
 the same setup by changing the link capacity of the bottleneck path (one of the two paths). 
 In Fig.~\ref{fig:short}, the top panel shows the variation in the bottleneck link capacity of the path. The bottom panel illustrates the reaction of model-based and DRL-based variants of MPTCP to the variation in link capacity 
 in terms of their data flow rates. The corresponding impacts of the path dynamics at the application level is demonstrated in terms of iteration time and unfairness observed at the worker's end. We have the following observations.

\subsubsection*{Performance issues over unseen conditions}
{\color{black} As illustrated in Fig.~\ref{fig:short}, we find that both DRL-based MPTCP methods and model-based MPTCP-LIA have severe performance problems under unanticipated time-varying circumstances (see the curves in Fig.~\ref{fig:short}). These efficiency problems include severe network under- and over-utilization and adversely impacts the DEL data parallelism.
A classic model-based MPTCP LIA~\cite{raiciu2011coupled} is simple to reason about and achieves a rather predictable efficiency under unknown circumstances but often fails to detect the change in path capacities in real-time. Despite convergence issues, it can be observed in Fig.~\ref{fig:short2} that DRL-based MPTCP~\cite{9444785} detects the changes in real-time and outperforms model-based MPTCP LIA~\cite{raiciu2011coupled} and TCP CUBIC in terms of fairness and iteration time perceived by the DEL applications. This is the marginal gain (see Fig.~\ref{fig:short2}) of MPTCP LIA  and DRL-based MPTCP in terms of iteration time and fairness when compared to using only TCP Cubic.}

\subsubsection*{Lack of driving force towards global equilibrium}
Besides, we find that DRL-based designs intelligently predict the time-varying path capacities (model-based TCP lacks this capability) however, they have convergence problems.
In particular, there
are the following two problems: i) converge to the different equilibrium points (e.g., see the difference between black and red curves from 10~s to 30~s in Fig.~\ref{fig:short}); ii) do not converge or converge slowly (from 0-10~s to 30-45~s). We anticipate that the lack of an agile process (e.g., the formulaic approach of model-based MPTCP LIA~\cite{raiciu2011coupled}) exacerbates the two problems and fails to promptly drive data transport dynamics and the DEL process towards an optimal equilibrium point. 

To overcome the 
aforementioned problems (convergence vs prediction trade-off), we develop a hybrid MPTCP (integrating successful features of robust model-based with the adaptive  DRL-based) to perform intelligent decisions of bandwidth allocations in real-time without undesirable fluctuations.

\subsection{Main Contributions}

Our major contributions in this paper can be outlined as follows.

\begin{itemize}
    \item [$\circ$] We model and develop an understanding of the network level causes for the fairness and efficiency in data parallelism of DEL.
\end{itemize}
   We find that multipath TCP has the capability to resolve the bandwidth allocation issues of DEL. However, solely learning-based MPTCP designs (e.g., DRL-MPTCP) emerging in the literature suffer from critical performance and convergence issues over unseen conditions. We demonstrate the under-utilization and over-utilization phases of the new clean slate DRL-based MPTCP designs with extensive emulations. We observe convergence issues of two types: i) converging to wrong equilibrium and ii) sluggish convergence (or sometimes may not converge). These new findings cater to the need for a more practical ML-based MPTCP design for DEL.
    
    \begin{itemize}
    \item [$\circ$] We develop a novel hybrid framework for analyzing and designing practicable ML-based MPTCP designs using successful features of both \textit{classic model-based} and \textit{modern DRL-based} MPTCP designs.
\end{itemize}
We implemented and evaluated the hybrid framework using a local testbed and demonstrated that the proposed hybrid MPTCP could consistently improve performance under various network settings.


\section{Hybrid MPTCP Design For DEL}

We observe that the bandwidth allocation and congestion control mechanisms for DEL require continuous control action as it is a sequential decision making problem that deals with how many bytes should be sent by the DEL workers to a network path at a particular point in time. Reinforcement learning is sequential and far-sighted (as in games). Deep neural networks are known to be successful in non-linear functional representations of complex problems, making them a better fit for learning the bandwidth allocation for DEL.

Nevertheless, such optimization of multipath bandwidth allocation for DEL is very slow and hard to implement with unconstrained approximates and large  (control action) state  spaces. We develop a hybrid approach where \textit{classic model-based MPTCP control loop is tightly coupled with the DRL agent using 
the actor-critic approach}. The gradient-based DRL algorithm can maintain an actor (function) and specify the policy by directly mapping states to a (specific) control action.  

{\color{black} Congestion control is a distributed algorithm that adapts
the \textit{congestion window (t)}, \textit{round trip time (t)} and \textit{loss(t)} in a closed loop. We omit the time $t$ in the expression for simplicity. Our Hybrid MPTCP Design employs a two-layer control loop mechanism.

Layer-1 is extended form of model-based MPTCP algorithm, an inner control loop,
{\color{black} where for every packet loss indication over path $i$ (and round trip time $\tau_i$)}, it performs multiplicative decrease 
and updates the congestion window $\omega_i'$ as $\omega_{i}' \leftarrow  \omega_i'/2$. 
In contrast to the multiplicative decrease for a packet loss, it employs continuous increase in the congestion window $\omega_i'$ for every successfully received Ack from path $i$ 
{\color{black} as $\omega_{i}' \leftarrow \omega_i' +\min\{1/\omega_i',\alpha(\omega_i')\}$  
\begin{equation}
\label{eqn:MPTCPwalia1}
\alpha(\omega_i') =  {\max_j\Big\{\frac{\omega_j'}{\tau_j^2}\Big\}}\Big/{\Big(\sum_j \frac{\omega_j'}{\tau_j}\Big)^2}; 
\end{equation}
}
Note that multiple data flows from multiple workers can share the same path. However, no multiple flows from a single worker can share the same path. One data flow rate from a worker $u$ with congestion window $\omega'_{i,u}$ over path $i$ is of rate $\theta_u^i=\omega'_{i,u}/\tau_i$. Therefore, the total rate from the worker $u$ over all paths, $\theta_u=\sum_i\omega'_{i,u}/\tau_i$ and therefore schedule is the probability vector, ${\bmath{h_u}}=\Big\{\frac{\theta^i_u}{\theta_u}\Big\}$.\\

Layer-2 consists of the DRL-agent-based outer loop, which runs as a daemon process to continuously monitor the network and traffic dynamics, compute the effective congestion window, $\hat\omega_i$, and schedule for all paths, and enforce them periodically so as to guarantee the target delay while harnessing transient network variations. The inner loop then adopts the enforced $\hat\omega_i$ (i.e., $\omega_i'\leftarrow \hat\omega_i$) and performs its fine adjustment
using~\eqref{eqn:MPTCPwalia1} till next reinforcements from the outer loop. As a result, the reinforced total rate from the worker $u$ over all paths can be interpreted as $\hat\theta_u=\sum_i\hat\omega_{i,u}/\tau_i$, and exploited for enforcing the new schedule, ${\bmath{\hat h_u}}=\Big\{\frac{\hat\theta^i_u}{\hat\theta_u}\Big\}$.\footnote{\color{black}Scheduling packets across multiple paths is a non-trivial challenging task. We frequently encounter a dilemma with the schedule vector. On the one hand, in order to harness bandwidth, we must split traffic across multiple paths; nevertheless, doing so impacts total latency since faster paths must wait for packets scheduled on the slower paths to be received. }}

The nonlinear approximation with DRL-agent design in Layer-2 is not desirable~\cite{li2019smartcc, xu2019experience}. In fact, such a nonlinear approximation is often unstable and may lead to difficulty in the convergence to the desired equilibrium. With relevant insights from Mnih \emph{et al.}~\cite{mnih2015human} and their findings for solving complex decision problems, we apply the logic of combining experience replay with convolutional neural network-enabled Q-learning for intelligent learning at outer loop for the efficient and fair bandwidth allocation of DEL. The proposed hybrid approach avoids instability and oscillations. It helps  adapt the underlying successful features of DRL~\cite{lillicrap2015continuous} in the continuous multipath bandwidth allocation and intelligent scheduling by harnessing the \textit{benefits of fairness and stability from model-based} MPTCP mechanisms.

\subsection{Fairness and Utility Maximization}
\subsubsection{Application Level Fairness and Utility}
Let $\mathcal U = \{1, 2, . . . , n\}$ represent the set of data flows from the workers. We consider the same length slotted decision interval operation to model the dynamics of the DEL process. At any point in time (integer multiple of the decision interval), a policy $\pi$ will generate a set ${\bmath{\theta}}$ = [$\theta_1, \theta_2, . . . , \theta_n$]
of expected total flow rates for the bandwidth allocation to workers over multiple paths.  From the application level perspective, the dynamic bandwidth allocation problem of DEL can be viewed as an objective to maximize the overall utility function $\mathbb U (\bmath{\theta})$ subject to the network path constraints as
\begin{eqnarray}
\max_\pi\mathbb U (\bmath{\theta})=\max_\pi\mathbb U (\theta_1, \theta_2, . . . , \theta_n)\; \mbox{s.t.}\; \bmath{\theta}\leq \bmath C.
\end{eqnarray}
{\color{black}Here, the path constraint means that the aggregate data flow rate $\theta_u^i$ from any worker over path $i$ cannot exceed its  effective maximum capacity $C_i$~\cite{pokhrel2017analytical}. Due to bandwidth sharing with other applications and the channel impairments, the capacity $C_i$ is time varying, the impact of which has been well studied in literature~\cite{pokhrel2017analytical}.}

With our DEL scenarios discussed earlier, we have the following two implications.

\begin{enumerate}
    \item [$\circ$] {\sc Improve Efficiency}. Considering data flow from a worker $u$ arriving at time $A_u$ and departing at $D_u$ to minimize
the computing and communication (upload/download latency) time of each worker by
\begin{eqnarray}
\min_\pi\Big(D_u-A_u\Big)\;  \forall u \in \mathcal U, \mbox{s.t.}\; \bmath{\theta}\leq \bmath C.
\label{eqn:3a}
\end{eqnarray}

The straggler, in synchronous schemes, impacts the momentum of the whole DEL system considerably. With~\eqref{eqn:3a}, we aim to minimize the overall completion time. For instance, an ill-suited policy always blocked the path with the smallest capacity, but an optimal approach would move the bottleneck from smaller capacity paths to larger capacity paths and reduce the overall completion time. 

\item [$\circ$] {\sc Enhance Fairness}. Considering the number of iterations of the data flow from a worker $u$ as $I(u)$ we minimize
the gap between the straggler and pioneers by
\begin{eqnarray}
\min_\pi\sum_u\Big( I(u)-\frac{\sum_{u'\in \mathcal U} I(u')}{|\mathcal U|}\Big)\;  \mbox{s.t.}\; \bmath{\theta}\leq \bmath C.
\label{eqn:3}
\end{eqnarray}
\end{enumerate}

\subsubsection{Flow Level Fairness and Utility}
By understanding the system performance through protocol and network parameters, we aim to utilize the limited network capacity better while ensuring each worker experiences satisfactory performance. Network bandwidth, instead, can be assigned amongst the contending workers based on several methods. To succeed in optimal bandwidth allocation, we intend to act towards efficiency and fairness at the data flow level. For instance, proportional and max-min fair bandwidth allocations~\cite{mazumdar1991fairness, kelly1998rate} are  widely accepted.
\begin{itemize}
\item [$\circ$]{{\sc Proportional Fairness}: A feasible bandwidth allocation [$\theta_1^*, \theta_2^*, . . . , \theta_n^*$] to the workers' set $\mathcal U$ is said to be proportionally fair iff for any other possible allocation of
[$\theta_1, \theta_2, . . . , \theta_n$], the 
following condition} 
holds true:
\begin{equation}
   \sum_{u\in\mathcal U} \frac{\theta_u-\theta_u^*}{\theta_u^*} \leq 0.
\label{eqn:prop}
\end{equation}

In the proportional fairness scheme, if the fraction of the change in the
data flow rate >$0$, there exists at least one other worker for which
the change is $<0$. We use proportional fairness in the performance evaluation of this 
paper and consider that the short term bandwidth allocation is equal for all data flows from/to the workers for jointly improving the performance and fairness at the application level.

\item[$\circ$] {\sc Max-min Fairness}: A feasible bandwidth allocation [$\theta_1^*, \theta_2^*, . . . , \theta_n^*$] to the workers' set $\mathcal U$ is said to be

\textit{max-min fair} iff $\theta_{u}^*$ for any $u\in \mathcal{U}$, allocated bandwidth cannot be increased without
decreasing $\theta_{u'}^*$ for some
${u'}\in {\mathcal U}$ for which
\begin{equation}
 {\theta_{u}^*}  \geq  {\theta^*_{u'}} .\nonumber
\end{equation}
Note that this scheme ensures that the need of workers is satisfied with
the minimum  possible resources.\\ 

\item[$\circ$] {\sc Utility function}: {\color{black}The utility of an individual worker in a DEL system can be taken either as the level of satisfaction of the worker or as a notion of network-level fairness~\cite{srikant2012mathematics}. Consider $U(\theta_u)$ the utility function of worker $u$ sending packets at 
a rate of $\theta_u$ packets/sec. The primary objective of the bandwidth allocation is to i.e., $\max\sum_{u\in U} U(\theta_u)$ $\forall u \in \mathcal U, \mbox{s.t.}\; \bmath{\theta}\leq \bmath C$. By using the log utility function, $U(\theta_u) = \log \theta_u$, we achieve  Eq.~\eqref{eqn:prop}~\cite{mazumdar1991fairness}). However, the proposed framework is flexible to handle several other notions of fairness, such as $\alpha$-optimal or $\alpha$-fair utility~\cite{srikant2012mathematics}.}
\end{itemize}

\subsection{DRL Learning Framework}
We consider a simple Markov Decision Process (MDP) with a discount factor $\gamma$ consisting of five-tuple  \[( \mathbb A(t), \mathbb S(t), \bmath R(t), \bmath T(t), \gamma),\] 
\begin{itemize}
    \item [$\circ$]$\mathbb A(t)$ is a finite set of actions, 
    \item [$\circ$] $\mathbb S(t)$ is a finite set of states,
    \item [$\circ$] $\bmath R(t)$ is a reward function, and
    \item [$\circ$] $\bmath T(t)$ is a transition function (computed using the state distribution of $\bmath P(t)$).
\end{itemize}
 For each state $s(t)\in\mathbb S(t)$, the DRL-agent actions $a(t)\in \mathbb A(t)$ to transition to a new state $s(t+1)$ with a reward of $\bmath R(a(t),s(t))$ is
mandated by the underlying probability distribution $\bmath P(s(t+1)|s(t),a(t))$.  

Our idea is to learn an optimal policy  for a DRL agent that determines for each state the appropriate action that 
the agent may take. The main objective of the DRL agent is to find the best fit policy that potentially maximizes the ultimate reward of the underlying bandwidth allocation for the DEL system. Similar to~\cite{winstein2013tcp}, we consider a 6-tuple state (\emph sending rates, throughputs, RTTs, change in windows, schedule and difference in RTTs) for a connection from a worker.\footnote{During the evaluation of the framework, we have found that adding more parameters increase the complexity of data sampling with negligible performance benefits.} 

The DRL-agent reinforces each workers' MPTCP connection and their subflows, which actions the ultimate changes (window increments, decrements and/or change in schedule) required for the allocation for each data flows and their subflows.

 We determine the optimal (greedy) policy, $\Pi^{\star}$ (based on $Q^{\star}$):
\begin{equation}
\Pi^{\star}(s)=\arg \max_{a'}Q^{\star}(s, a),
\label{eqn:greedy}
\end{equation}where $Q^{\star}$, for a given state-action pair, is predicted by the Bellman equation~\cite[pp. 2]{hester2018deep}:
\begin{equation}
Q^{\star}(s,a)=\underset {s'\sim P} {\mathbb E}\big[R(s, a, s')+\gamma \max_{a'}Q^{\star}(s', a') \big].
\end{equation}

As a result, the DRL agent can approximate its optimal $\Pi^{\star}$ simply by using the $Q^{\star}$ approximator in~\eqref{eqn:greedy}. 
This can be perceived as a type of prioritized experience-driven approach, which determines a mechanism to move steadily towards a comparatively better performance  even with small observations and a limited learning period~\cite{lillicrap2015continuous, hester2018deep}.
 
\subsection{Hybrid MPTCP: Working Mechanism}
\label{sec:design}
\label{sec:abstractview}
\begin{figure}[t]
\centering
\includegraphics[width= 3.1865 in]{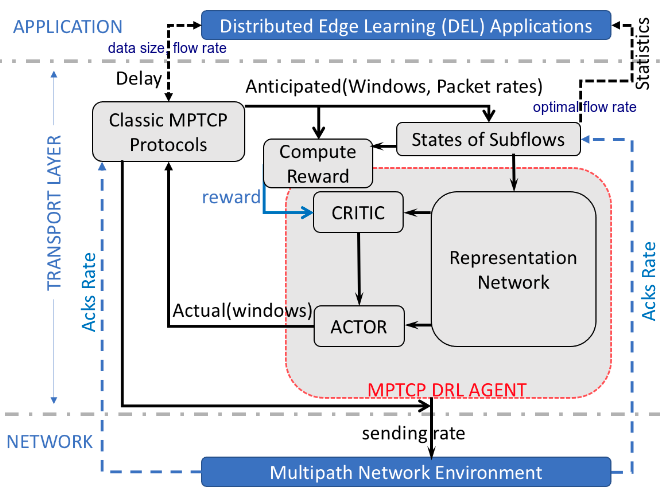}
\caption{\rm High-level view of the components of Hybrid MPTCP and their interrelationship. The components inside the dashed region are the outer loop (forming the DRL-agent). The DRL agent takes  model-based windows, schedule and sending rates as inputs and generate actual windows and data flow rates as outputs. A component on the left top box acts as an inner loop (a classic model-based MPTCP module with Ack-based window evolution), which provides inputs to the DRL agent. Their interactions with the DEL and application layer constraints are via efficiency and fairness measures, and the data flow statistics are provided as output. The underlying network layer sees packets sending rate as input and provides Acks rate as outputs. \label{fig:MPTCP-blocks}}
\vspace{-5 mm}
\end{figure}

In this subsection, we elaborate on the details of the interaction between different components of our \emph{DRL-enabled hybrid MPTCP} for bandwidth allocation and data parallelism of DEL under different circumstances. At a higher conceptual  level, the components of the proposed hybrid MPTCP and their interactions is shown as in Fig.~\ref{fig:MPTCP-blocks}.

Fig.~\ref{fig:MPTCP-blocks} explains the interrelationship with the modules inside the dashed region as the outer loop (DRL-agent) that takes  windows and data flow rates as inputs, and generates control actions (actual windows and effective allocation rates) as outputs. On the left rectangle is another module that acts as an inner loop (is a classic MPTCP mechanism with Ack based window evolution at the kernel adopted from~\eqref{eqn:MPTCPwalia1}), which provides new schedule, data flow rates and congestion windows to the DRL-agent.
{\color{black}
Their interaction with the DEL application layer constraints is through utility, efficiency and fairness constraints (recall (3) and (4)) and optimal data flow rate and traffic statistics (outputs). The underlying network layer sees the sending rate of packets as an input and provides Acks as outputs.}

Algorithms explaining the steps  of the DRL-agent (actor-critic training, representation network and state {analysis} of the subflows, see Fig.~\ref{fig:MPTCP-blocks}) are provided in subsequent subsections. The proposed hybrid MPTCP interacts with the multiple paths, model-based MPTCP component and the DEL applications to update the state information $\mathbb S(t)=\{s(t, i)\}$.
With slight abuse of notation, $i$ indicates a subflow of a DEL data flow from a worker on 
path $i$. 

At the end of a slotted reinforcement time frame $t$, 
reward $\bmath R(t)$  can be computed as the aggregation of individual utilities of all active subflows (based on previous actions, it can be estimated by using the utility formulation of model-based MPTCP) by using states of the subflows, which is supplied to the critic network (Sec.~\ref{sec:accrt}). The computation of reward is based on the representation network (Sec.~\ref{sec:netrep}), learned by the overall process and the current state of subflows $\mathbb S(t)$ (e.g., $\bmath R(t)$ is computed by using final states $\mathbb F(t)$ and action $\bmath a(t)$).  

The DRL-agent supplies actions to the extended version\footnote{This is an extension to the implementation of model-based MPTCP which enforced $\omega_i$s for its subflows as actioned by the DRL-agent.} of the MPTCP source (once per every slotted time frame) to:
\begin{itemize}
    \item renew the size of the windows ($\mathbb W(t+1)$), as per perceived RTTs ($\bar{\bmath\tau}=\{\bar\tau_i\}$) of active subflows via the Acks rate and determined by the DRL-agent; and 
    \item {\color{black}adjust the schedule (${\bmath{h_u}}\leftarrow\bmath{\hat h_u}$) by using~\eqref{eqn:3a} and considering application level delay. This mechanism intelligently determines the routing probability of the packets to the subflows (cf. Fig.~\ref{fig:MPTCP-blocks}).}
\end{itemize}

  It is worth observing that instead of using the actions straightforward, a policy gradient formulation with the agent's representation is adopted for all action values. This predicts the required gradients. Then, the proposed hybrid MPTCP framework directly executes
the action (via scheduling and bandwidth allocation at
the MPTCP source in the kernel) by monitoring the
reward $R(t)$. 

\subsection{Representation Network}
\label{sec:netrep}
The representation network maintains a (representation) vector for continuously monitoring the states of hybrid MPTCP subflows. As shown in Fig.~\ref{fig:MPTCP-blocks}, it takes $\mathbb S(t)=\{s(t,i)\}$ as inputs. 

The main difficulty here is to deal with the representation network when the number of data flows, and their subflows are time-varying (joining and leaving the paths by the workers with the same MPTCP flow using different combinations of paths over time), which appears to be very common in next-generation wireless networks. Therefore, the usual fixed input size of the Deep Neural Network (DNN) approach is not suitable. Rather, we need to employ a well-designed LSTM with an intrinsic capability to accurately track the variable number of inputs to quantify the dynamics of time-varying subflows. Such a choice for representation could also be based on the corresponding availability and capacity of the network path(s). We implement the sequential approach in order
to manage the states of the data flows and their subflows into the LSTM one after another, as it will be more practical for learning the representation network in a linear fashion.

The output of the representation network is the final state $\bmath f^N(t)$ ($\mathbb F(t)$ represents all the subflows of a data flow from the same worker), which is used by the Actor-Critic for training. Our approach trains the triad Actor, Critic and Representation (ACR) jointly with backward propagation. For the joint training, we maintained a separate inner connection between four modules, viz. \emph{state of subflows $\rightarrow$ network representation $\rightarrow$ actor-critic training}, which maintains synchronization among the three networks (as shown by the closed-loop in Fig.~\ref{fig:MPTCP-blocks}). Our analysis and implementation
have remarked that the ACR networks' joint training 
can enhance performance substantially compared to training them separately.


\subsection{Training the ACR Networks}
\label{sec:accrt}
The actor-network  is a connected mesh of
LSTM. It consists of two hidden layers 
of 128 neurons each. The critic network is the same as the actor with an additional (output) layer of a linear single layer neuron. In the hidden layers, we use rectified
linear function, but in the output layer, we use the hyperbolic tangent function for dual activation. We use Adam optimizer ({learning rates are set to
{\color{black} 0:0009 and 0:009}, and discount factor $\gamma= 0.97$}) for training the ACR networks, which is constructed by using higher-level API to TensorFlow\footnote{\emph{http://tflearn.org}} of the neural networks.


\subsection{Computing Reward}
\label{sec:accrt}
The computation of reward in this hybrid MPTCP framework is motivated by the network utility maximization concept of MPTCP~\cite{peng2016multipath}; in particular, 
the flow level reward here is the sum of the utilities of all active subflows belonging to the data flow from a worker,
which is given by 
\begin{eqnarray}
R(t):=\sum_i U_i(t),
\end{eqnarray} 
where $U_i(t)$ is the utility of subflow $i$ at time $t$.

{\color{black}The proposed hybrid MPTCP framework adopts flexible plugins. We can adopt desired utility functions (e.g., fairness-based or loss, delay, and throughput-based) with required scheduling policies. In our evaluation, we apply the well-known proportional fairness approach for all subflows (active paths). In particular, we aim to maximize the sum of functions, $U_i(t)=\log \theta_u^i(t)$, over all paths, where $\theta_u^i(t)$ is the short-term average rates over path $i$ in the previous iteration. For delay consideration, our scheduling policy implements batch scheduling of packets to the subflows with a rate proportional to their short-term average rates.}
\begin{algorithm}[!t]
\caption{Training ACR Networks}
\label{Training algorithm}
\begin{algorithmic}[1]
\Procedure{\color{red} Training ACR} {}\\
Inputs: Schedule and Rates from classic MPTCP\\
Inputs-based sampling of ($\mathbb S_i, \bmath{{a}}_i, \bmath R_i, \mathbb S_{i+1} $) as transitions from replay buffer\\
Computing final state $\bmath f^N_{i+1}$ by using $\mathbb{\hat{N}}(\mathbb S_{i+1})$\\
Evaluating Critic $\bmath Q(.)$, ${\footnotesize y_i= \bmath R_i+\gamma \bmath{\hat{Q}}(\bmath f^N_{i+1}, \mathbb{\hat{E}}(\bmath f^N_{i+1}) )}$\\
Parameters fo Critic $\min(loss)$, ${\footnotesize 1/k\sum_i^k(y_i-\bmath Q(\bmath{{a}}_i, \bmath {f^N_{i}} ))^2}$\\
Evaluating Gradient from Critic: $\bigtriangledown_a\bmath Q(\bmath{{a}}_i, \bmath {f^N_{i} })$\\
Actor parameters updating, ${\footnotesize 1/k\sum_i^k\bigtriangledown_a\bmath Q(\bmath{{a}}_i, \bmath {f^N_{i}}). \bigtriangledown_{k}\mathbb E(\bmath{f^N_{i}})}$\\
Actor-based gradient computing  $\bigtriangledown_k\mathbb E(\bmath{f^N_{i}} )$\\
Updating Representation Network (with gradients),\ \ \ \ \  \ \;\;\;\;
${\footnotesize {1/k\sum_i^k\bigtriangledown_{\bmath{a}}\bmath Q(\bmath{a}_i, \bmath {f^N_{i}}). \bigtriangledown_k\mathbb E(\bmath {f^N_{i}}).\bigtriangledown_{\bmath R_i}\bmath R(\mathbb{S}_{i+1}}) }$
 {\color{red}\EndProcedure}
\end{algorithmic}
\end{algorithm}

\subsection{Training and Learning Algorithms}
\label{sec:algo12}
ACR  networks training and DRL-agent learning processes are shown by Algorithms~\ref{Training algorithm} and~\ref{MPTCPalgorithm},
respectively. The training process in Alg.~\ref{Training algorithm} consists of evaluating critics with loss parameters and then computing actor-based gradients for updating representation networks. Alg.~\ref{MPTCPalgorithm} is 
a daemon process that always runs as a background and starts with initializing the parameters of the ACR networks $\mathbb N(.)$,
$\bmath Q(.)$, and  $\mathbb E(.)$. As a result, the DRL-agent learning process runs all the time, listening for regular queries from the classic MPTCP source module (implemented by using the system calls at the kernel).

When training an amateur DRL-agent,  we combine correlated noise by employing the stochastic Ornstein-Uhlenbeck stochastic process to facilitate exploration~\cite{lillicrap2015continuous}. This is to regulate actions dynamically. Such an intrinsic integration of experience replay with the exploration benefits the DRL agent by efficiently adjusting subflow bandwidth allocation based on the desired degree of efficiency and fairness for the data parallelism of DEL.   

For ensuring convergence in the learning process (see step 3, Alg.~\ref{MPTCPalgorithm}) , target ACR networks $\mathbb{\hat{N}}(.)$, $\mathbb{\hat{E}}(.)$, and $\bmath{\hat{Q}}(.)$ are used for replicating the structure of the ACR.

These updates of these target ACR networks evolve at a slower time scale as they are based on very small control parameters (0.005, in steps 15-17, Alg.~\ref{MPTCPalgorithm})
such that the target parameters are adapted slowly in each iteration for stability. The target ACR networks, by the proposed design, need to update slowly for stability. We have a daemon process, and our agent runs all the time, tracking for the periodic updates inquired by the MPTCP source the inputs from the DEL system. As shown in step 8 of Alg.~\ref{MPTCPalgorithm}, the $\mathbb{F}(.)$ and allocation is computed with $\mathbb{N}(.)$, while actions for
subflows $\hat{\bmath a}(t)$ depend on $\mathbb{E}(.)$ (indicated by step 9 in  Alg.~\ref{MPTCPalgorithm}). 

All of the created samples of the transitions  are backed up into the memory for randomly sampling them later during ACR training
(\emph{$\mathbb N$, $\mathbb E$, $\bmath Q$}) jointly by sampling (step 2, Alg.~\ref{Training algorithm}). The critic network as shown in step 5 (Alg.~\ref{Training algorithm}) is a \emph{deep Q-Learning}; 
its updates are based on minimizing the squared error and Bellman formulation (see step
5-6 of Alg.~\ref{Training algorithm}). With the
Bellman equation,  Q-function processes  actions ($\bmath{a}(t)$) and states ($\mathbb S(t)$). For continuous dynamics tracking of the training process, the parameters of the ACR evolved with the gradients based on chain rule (see details in~\cite[Eqn. (6)]{lillicrap2015continuous}). This is feasible with the  sampling process consisting of $s$ samples as indicated from step 6 to step 10 of Alg.~\ref{Training algorithm}. 
\begin{algorithm}[t]
\caption{DRL-agent Learning Algorithm}
\label{MPTCPalgorithm}
\begin{algorithmic}[1]
\Procedure{\color{blue} Daemon} {}\\{
Process $\mathbb a$ for Ornstein-Uhlenbeck exploration\\
 Actor $\mathbb E(.) $,Representation $\mathbb N(.)$ and Critic $\bmath Q(.)$\\
Target \{ Actor $\mathbb{\hat{E}}(.) $\;Representation\; $\mathbb{\hat{N}}(.)$\; Critic $\bmath{\hat{Q}}(.)$\}\\
Inputs: Schedule and Rates from classic MPTCP
\Procedure{\color{violet}Bandwidth Allocation} {}\\
\hspace{4 mm}\textbf{While} {(Schedule || Data Flow Rate change)} \textbf{do}\\
\hspace{5 mm}Final state $\mathbb F(.)$ and schedule computation with $\mathbb N(.)$\\
\hspace{5 mm}Target updating: $\bmath{\hat{a}}(t)$ using Actor $\mathbb E(\bmath f^N(t))$\\
\hspace{5 mm}Action generation for classic MPTCP, $\bmath{{a}}(t)$: ($\bmath{\hat{a}}(t)$, $\mathbb a$)\\
\hspace{5 mm} Transitions back-up:  ($\bmath{{a}}(t), \mathbb S(t), \bmath R(t), \mathbb S(t+1) $)

\Procedure{\color{red} Continuous Training} {}\\
 \hspace{8 mm} Algo.~1  ({\sc Training ACR})
{\color{red} \EndProcedure}
\Procedure{\color{blue} Target ACR Network} {}\\
\hspace{18 mm}$\mathbb{\hat{E}}(.):=.005\mathbb{{E}}(.)+.995\mathbb{\hat{E}}(.)$\\
\hspace{18 mm}$\mathbb{\hat{N}}(.):=.005\mathbb{{N}}(.)+.995\mathbb{\hat{N}}(.)$\\
\hspace{18 mm}$\bmath{\hat{Q}}(.):=.005\mathbb{{C}}(.)+.995\bmath{\hat{Q}}(.)$
{\color{blue} \EndProcedure}\\
\hspace{4 mm}\textbf{endWhile}
\hspace{18 mm}{\color{violet}\EndProcedure}}
{\color{blue}\EndProcedure}
\end{algorithmic}
\end{algorithm}

\section{Implementation and Performance Evaluation}
\label{sec:trainingandeval}

{\color{black}We have programmed our Hybrid MPTCP framework by reusing the Linux Kernel implementation of MPTCP v0.93. For convenience and ease of implementation and future deployment, we use a dedicated DRL agent in the user-space which interacts with the DEL application and the Kernel to adjust the windows and schedule  of the connection. Note that in LIA~\cite{raiciu2011coupled}, BALIA~\cite{peng2016multipath} and others, the later adjustments are also done in the Kernel for each connection, but according to fixed rules.   

For tractability and interoperability with the existing MPTCP implementation, we have developed our DRL agent as a plug and play user-space module based on the Linux's specification, while embedding the final trained model into the Kernel may improve the system performance. Every MPTCP connection reserves a memory for all of their subflows, such that DRL Kernel handler (and the DRL agent) fetch the windows and rates of all subflows from their Kernel (memory) space periodically. The DRL-agent interacts with DEL application, generates an action and enforces the action via the Kernel handler for dynamically updating the values of the windows and schedule for the connection.}
 

\subsection{DRL-Agent Training}
We have experimented with training  analyzing and evaluating the performance of the hybrid MPTCP under several network scenarios using a trace-based network emulator~\cite{netravali2015mahimahi}
to handle more easily the output of different MPTCP protocols in a reproducible setting. The range of emulated environment of the training are: i) link bandwidths (4 to 128 Mbps), ii) round trip times ($3-300$ ms) and iii) bottleneck buffer size ($20 -500$ TCP packets of 1500 bytes). To confirm trained models, rigorous training for both hybrid MPTCP and DRL-MPTCP~\cite{pokhrel2020multipath} are repeated \textit{three times} each with the same data sets. The parameters used in training are shown in Table~\ref{tab:params}. During training sessions, we measure the rewards for the Hybrid and DRL-based MPTCP. We average out the rewards for each scheme of all actors (called \textit{learning score}) for all training sessions to aggregate the outcomes of all training sessions.

\begin{figure}[t]
\centering 
\includegraphics[width=2.025 in]{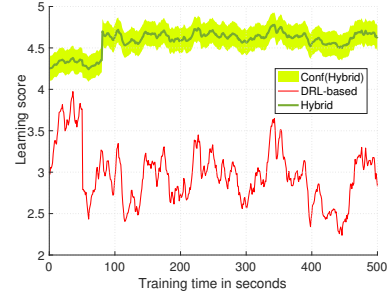}
\caption{\rm \textbf{Training I}: One snapshot of the worst training curves of Hybrid and DRL-MPTCP under the same setting. \label{fig:fig4}} 
\vspace{-5 mm}
\end{figure}

\begin{figure}[t]
\centering 
\includegraphics[width=2.0802 in]{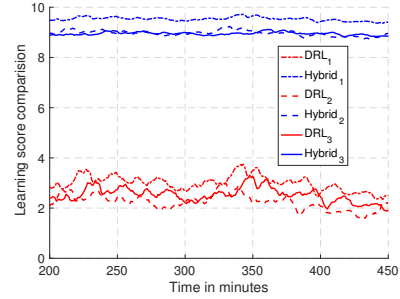}
\caption{\rm \textbf{Training II}: Moving average training curves of Hybrid and DRL-MPTCP under the same setting for all three sessions (each training session is repeated four times to guarantee that the learned model is not impacted by the undesirable random weights).
\label{fig:fig5}} 
\vspace{-5 mm}
\end{figure}

The summarized results of all training sessions for both Hybrid and DRL-based MPTCP, as observed in one session (worst case) and average overall three sessions, are shown in Figs.~\ref{fig:fig4} and~\ref{fig:fig5},
respectively.

\begin{table}[t]
\centering
\caption{Training Parameters. \label{tab:params}}
\begin{tabular}{ l | l| l |l }\hline\hline
\bf Parameters & \bf Value & \bf Parameters & \bf Value \\ \hline
Learning rate (Actor) &$0.0009$&
Learning rate (Critic)&$0.009$ \\
Optimization &ADAM &
Coefficients of Target&$0.005$ \\
Discount Factor &$0.97$ &
Samples/session &$30,000$ \\
$\mathbb a$ exploration noise& $\mathcal N(0,0.2)$ &
 Size N& 2024\\\hline\hline
\end{tabular}
\vspace{-2mm}
\end{table}

By comparing the performance of Hybrid MPTCP with DRL-MPTCP, in Figs.~\ref{fig:fig4} and~\ref{fig:fig5}, we have found that the Hybrid MPTCP outperforms DRL-MPTCP even during the training. It is worth noting that the inner loop consisting of the classic model-based MPTCP control which significantly contracts the action space of the DRL-agent 
of the proposed hybrid MPTCP, and accelerates the training dramatically compared to DRL-MPTCP. In particular, as 
observed in Fig~\ref{fig:fig5}, the learning score of Hybrid MPTCP is  approximately three times more compared to the DRL-MPTCP.

\subsection{Evaluating Hybrid MPTCP}
Considering the challenges of data parallelism in DEL, we evaluate the proposed hybrid DRL-MPTCP algorithm and compare its performance with the existing MPTCP versions (for instance, LIA~\cite{raiciu2011coupled} and DRL-MPTCP~\cite{pokhrel2020multipath}). We have utilized extended Linux implementation of MPTCP v0.93.\footnote{ \emph{https://www.multipath-tcp.org}}
 
 As workers and a server, our testbed consists of seven laptops.
All six laptops are connected via a switch (Gigabit), which consists of two independent interfaces (Gigabit Ethernet), creating two different connection paths for each worker, constantly uploading/downloading big files from the server.  
 
 For tractability, along the lines of~\cite{peng2016multipath}, each data flow between the server and laptop  consists of two subflows in our experiments. In order to experiment not only with varying bandwidth but also to evaluate the impact of varied asymmetric loss and delays, we have adopted and experimented with Linux data control features.\footnote{\emph{NetEm:{https://wiki.linuxfoundation.org/networking/netem}}} In the controlled experimental setting, we fixed 12 paths between the laptops and the server (each data flow having two subflows).
Our results are based on different dynamics of traffic data when uploading/downloading data of varying sizes (3MB to 900MB).
The data retrieved from TCP dumps reflect an aggregate average of about fifteen hundred runs and values in the graphs (values in the plots represent the overall mean value of all runs unless stated otherwise).

Our main results are discussed as follows.

\subsubsection{Experiment 1} {\color{black} We  first analyze the impact of time-varying bandwidth of the bottleneck path. We have emulated distributed edge learning (shown in Fig.~7) which involves 500 model files ($600$~MB each) uploads/downloads. We  vary 
the capacity of the bottleneck path (as shown in Fig.~\ref{fig:short} top panel) over time dynamically while keeping the capacity of other paths fixed at 8~Mbps. By setting equal delay ($50$ms) and equal loss of $3\%$ along both paths we conduct  five hundred uploading/downloading using models of $600$~MB. After performing different runs for \emph{Hybrid, DRL-based MPTCP and MPTCP-LIA} over the same setting, we observe the impact of time-varying bottleneck capacity
on data flow rate perceived by the workers. Figs.~\ref{fig:short} and~\ref{fig:fig6bandwidth} show
the reaction of Hybrid, LIA and DRL-MPTCP.}

In terms of data sending rates, for the same size data flows, our hybrid MPTCP outperforms the model-based MPTCP and DRL-based algorithms. 
 In the first case, with decreasing capacity, 
 we can observe in Fig.~\ref{fig:fig6bandwidth} that the joint effect of model-based MPTCP and DRL agent in our hybrid MPTCP (smoothing the transition with early prediction). Similar behaviours 
 are observed in the second case while switching from 
 a lower capacity to a higher one at the time stamp of \textit{30-35} secs.

 We believe that this finding is due to the joint effect of the hybrid (model-based and DRL-based) policy adopted in this paper. However, a complete theoretical understanding of the dynamics in the DRL agent and how such a hybrid learning-based approach works well in practice is still an open research  question and requires further investigation in future.

\subsubsection{Experiments 2 and 3}
We conduct two different experiments, 
namely, Experiment 2 for synchronous data parallelism and Experiment 3  for the asynchronous data parallelism cases under the same setting, when DEL data flows from working using MPTCP are competing with other data flows using TCP CUBIC connections. By extending the setup of Experiment 1, we adjust capacities from 20 Mbps to 50 Mbps (for both paths) and examine the aggregate  performance with 
six data flows using MPTCP and  sharing one of the paths with rival TCP CUBIC based data flows. This is conducted repeatedly with all the three variants: Hybrid, DRL-MPTCP, and 
MPTCP LIA. To evaluate the practicality of the three variants of MPTCP, as shown in Figs.~\ref{fig:fig7} and~\ref{fig:fig81}, we compared the fluctuation in their throughputs (average values of the changes in throughput between two consecutive timings). We have the following two interesting findings:  
\begin{figure}[t]
\centering 
\includegraphics[width=3.02825 in]{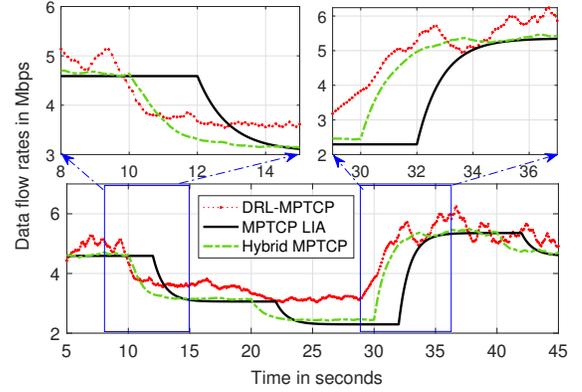}
\caption{\rm {\color{black} \textbf{Experiment 1}: We  vary the capacity of one of the links (as shown in Figure~\ref{fig:short} top panel) over time dynamically and fixed another path at 8~Mbps). We set delay ($50$ms) and $3\%$ path loss equal for both paths while continuously uploading/downloading replicas of $600$MB  file. We perform different runs for \emph{DRL-based MPTCP and MPTCP-LIA} with 
the same setup by changing the link capacity of only one of the two paths to observe the impact of time-varying capacity on data sending rate attained by the hybrid and DRL-MPTCP connections. \label{fig:fig6bandwidth}} }
\vspace{-4 mm}
\end{figure}
\begin{figure}[t]
\centering 
\includegraphics[width=2.67825 in]{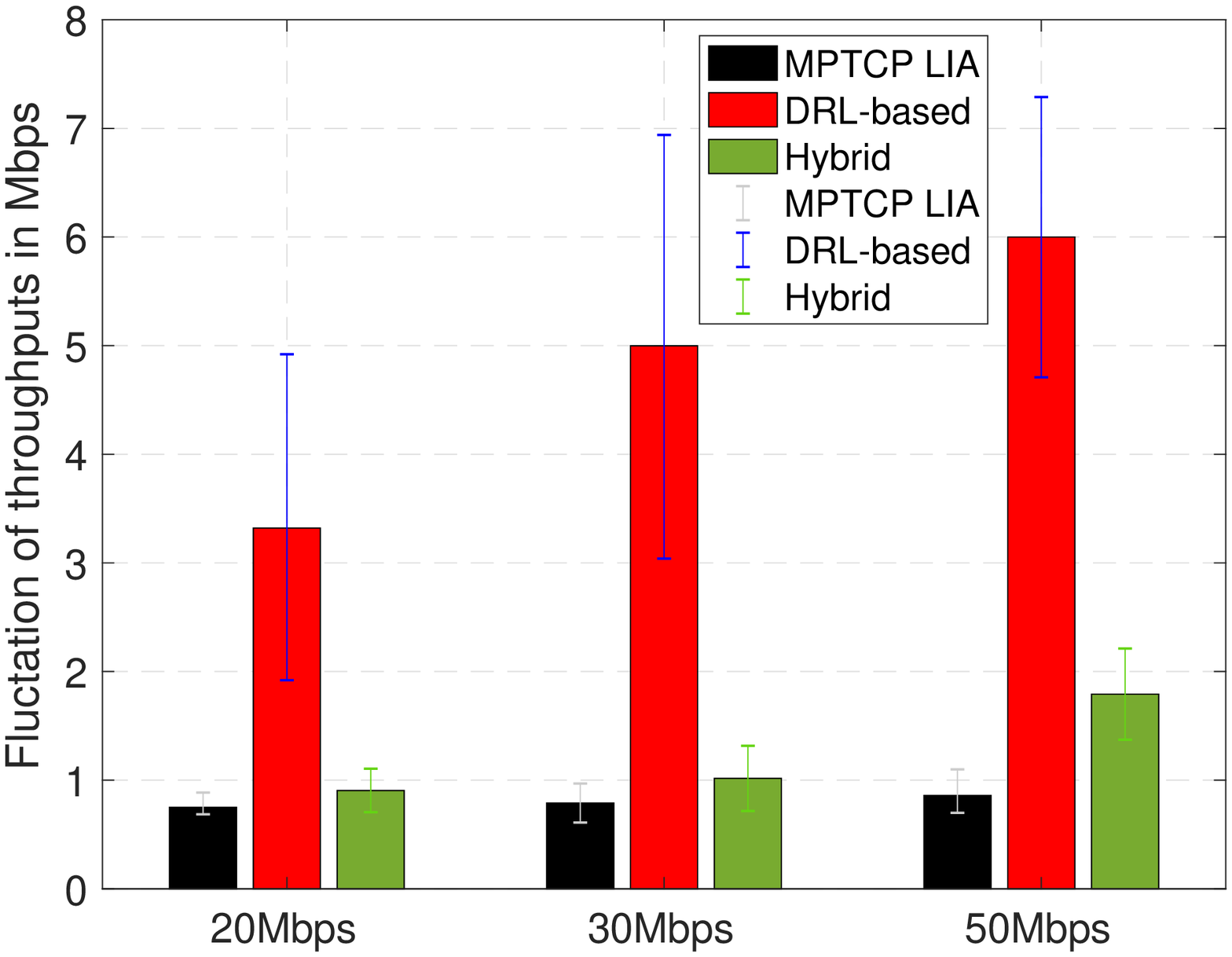}
\caption{\rm \textbf{Experiment 2}: Comparison of data flow rate fluctuation between the three algorithms under Synchronous Data Parallelism Schemes. Data flows using MPTCP are competing with other data flows using TCP CUBIC connections. Using the extended set up of Experiment 1, we adjust path capacities from 20 to 50 Mbps and experiment with six data flows using MPTCP connections  sharing one of the paths with other data flows using TCP CUBIC. The experiment is conducted repeatedly for all three variants: Hybrid, 
DRL-MPTCP, and LIA.\label{fig:fig7}} 
\vspace{-4 mm}
\end{figure}\\
\begin{figure}[t]
\centering 
\includegraphics[width=2.678925 in]{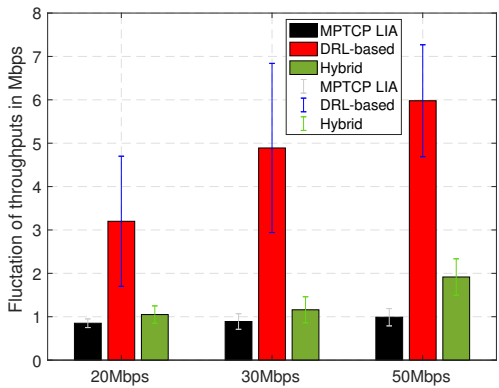}
\caption{\rm \textbf{Experiment 3}: Comparison of data flow rate fluctuation between the three algorithms under Asynchronous Data Parallelism Schemes. Data flows using MPTCP are competing with other data flows using TCP CUBIC connections. Using the extended setup of Experiment 1, we adjust path capacities from 20 to 50 Mbps, and experiment with six data flows using MPTCP connections  sharing one of the paths with data flows using TCP CUBIC. The experiment is conducted repeatedly for all three variants: Hybrid, 
DRL-MPTCP, and LIA.\label{fig:fig81}} 
\vspace{-4 mm}
\end{figure}

\begin{itemize}
    \item As shown in Figs.~\ref{fig:fig7} and~\ref{fig:fig81}, the rate of fluctuations of throughputs of the Hybrid MPTCP is close to MPTCP LIA, demonstrating the practicability of the algorithms. {DRL-MPTCP} perceived the highest fluctuation (Figs.~\ref{fig:fig7} and~\ref{fig:fig81}, questioning its practicability).  \item All three MPTCP algorithms have little difference between asynchronous and synchronous in throughput fluctuations, however, both MPTCP LIA and DRL-based fall behind the hybrid scheme in terms of utilizing the varying bandwidths dynamically (see Figs.~\ref{fig:fig7} and~\ref{fig:fig81}).
\end{itemize}

It is worth noting that we observe Hybrid MPTCP introducing the iterations at the cost of hybrid objective function under both parallel schemes. As a result, under asynchronous parallel schemes, its actions seem
to be consistent in the same iteration for both schemes, making its fluctuation lower than DRL-based MPTCP.

\subsubsection{Experiment 4}

With the settings of Experiment 3, we  conduct experiments with six data flows from workers using MPTCP connections sharing one path with the data flows from workers using TCP CUBIC. We evaluate the throughput behaviour of the straggler over time.  We repeat the same experiment by using the three variants, DRL-based, MPTCP LIA, and Hybrid.

\begin{figure}[t]
\centering 
\includegraphics[width=2.67925 in]{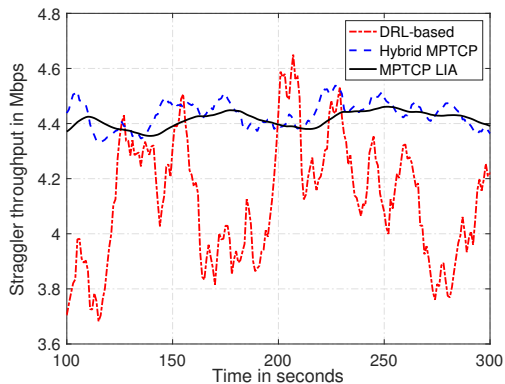}
\caption{\rm \textbf{Experiment 4}: Perceived throughput behaviour of the slowest worker (straggler) over time with the three different algorithms. DEL data flows from workers using MPTCP are competing with other data flows using TCP CUBIC. Using the setup of Experiment 3, we conduct repeated runs for all the three variants: DRL-based, MPTCP LIA, and Hybrid.\label{fig:fig91}} 
\vspace{-4 mm}
\end{figure}

 Fig.~\ref{fig:fig91} illustrates the moving average throughput comparisons for all three algorithms, where they are competing with a single path TCP flow.
 Since DRL-based is fully learning-based, its fluctuations are the highest. The fluctuation of MPTCP LIA is the smallest. The case for the hybrid is very similar to MPTCP LIA, showing its improved stability.
 
\subsubsection{Experiment 5}

\begin{figure}[t]
\centering 
\includegraphics[width=2.467825 in]{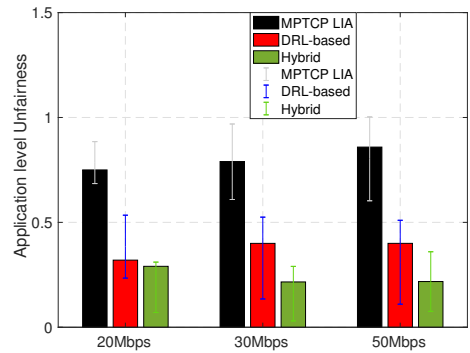}
\caption{\rm \textbf{Experiment 5}: Comparison of application-level unfairness when data flows from workers using MPTCP connections are competing with TCP CUBIC based data flows in one of the paths. With the same setup of Experiment 4, we repeat experiments using three MPTCP variants: Hybrid, DRL-based and LIA.\label{fig:fig101}} 
\vspace{-4 mm}
\end{figure}

To evaluate the iteration time and unfairness, under asynchronous data parallelism schemes,  we evaluate average iteration time and unfairness (as the average standard deviation over workers iterations) using the setup of Experiment 3, where we  use six data flows from working using MPTCP connections  sharing one of the paths with other data flows using TCP CIBIC. As shown in Fig.~\ref{fig:fig101}, the  fairness of MPTCP LIA for data parallelism  is the lowest, while DRL-based and Hybrid attain a high degree of fairness (very close to each other). In particular, we observe that hybrid MPTCP, most of the time, ensure that the difference between the pioneer and the straggler is very narrow and, therefore, performs even better than the DRL-MPTCP. This is due to the coupling of our hybrid MPTCP with application-level constraints.

\begin{figure}[t]
\centering 
\includegraphics[width=2.467825 in]{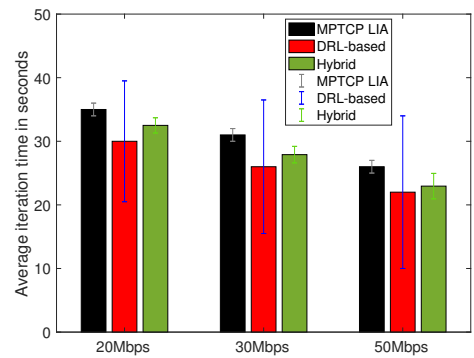}
\caption{\rm \textbf{Experiment 5}: DEL data flows with multipath TCP and single-path TCP competing with each other. With the setup of Experiment 4, we  use five data flows over MPTCP, which are sharing a bottleneck path with rival five data flows over TCP. With repeated experiments, observe the average DEL iteration time perceived by the workers in all three different versions of MPTCP (DRL-based, MPTCP LIA,
and Hybrid MPTCP).\label{fig:fig102}} 
\vspace{-4 mm}
\end{figure}

As illustrated in Fig.~\ref{fig:fig102}, we also analyzed and compared the difference in mean iteration time of the workers when using the three variants (DRL-based, MPTCP LIA, Hybrid MPTCP). With the same settings of Experiment 3, we vary link capacities from 20 Mbps to 50 Mbps for both paths and examine the aggregate  performance with six data flows from workers using MPTCP competing with rival data flows using TCP CUBIC over one of the paths. We found that compared to MPTCP LIA, our proposed hybrid MPTCP reduced the average iteration time considerably. However, it is a little bit higher than that of DRL-based. Furthermore, it is worth noting that the gap between DRL-based and Hybrid MPTCP decreases with an increase in link bandwidth; a complete understanding of this trade-off requires further investigations and will be investigated in future. 


 \begin{figure}[b]
\centering 
\includegraphics[width=2.5799825 in]{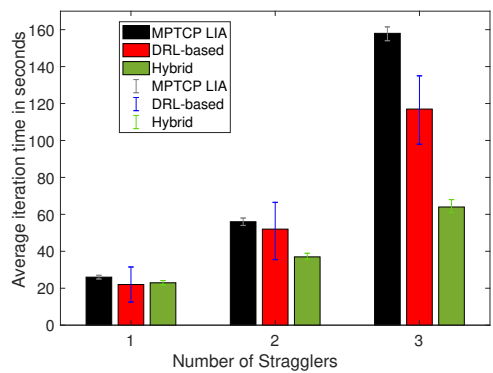}
\caption{\rm {\color{black}\textbf{Experiment 6}: DEL data flows with multipath TCP and single-path TCP competing with each other. With the setup of Experiment 5, we  fixed the bandwidth to 50 Mbps, and change the number of stragglers. We repeat the same experiment to observe the average DEL iteration time perceived by the workers in all three different versions of MPTCP (DRL-based, MPTCP LIA,
and Hybrid MPTCP). \label{fig:fig102s}} }
\vspace{-4 mm}
\end{figure}
\begin{figure}[t]
    \centering
    \includegraphics[scale=0.26345]{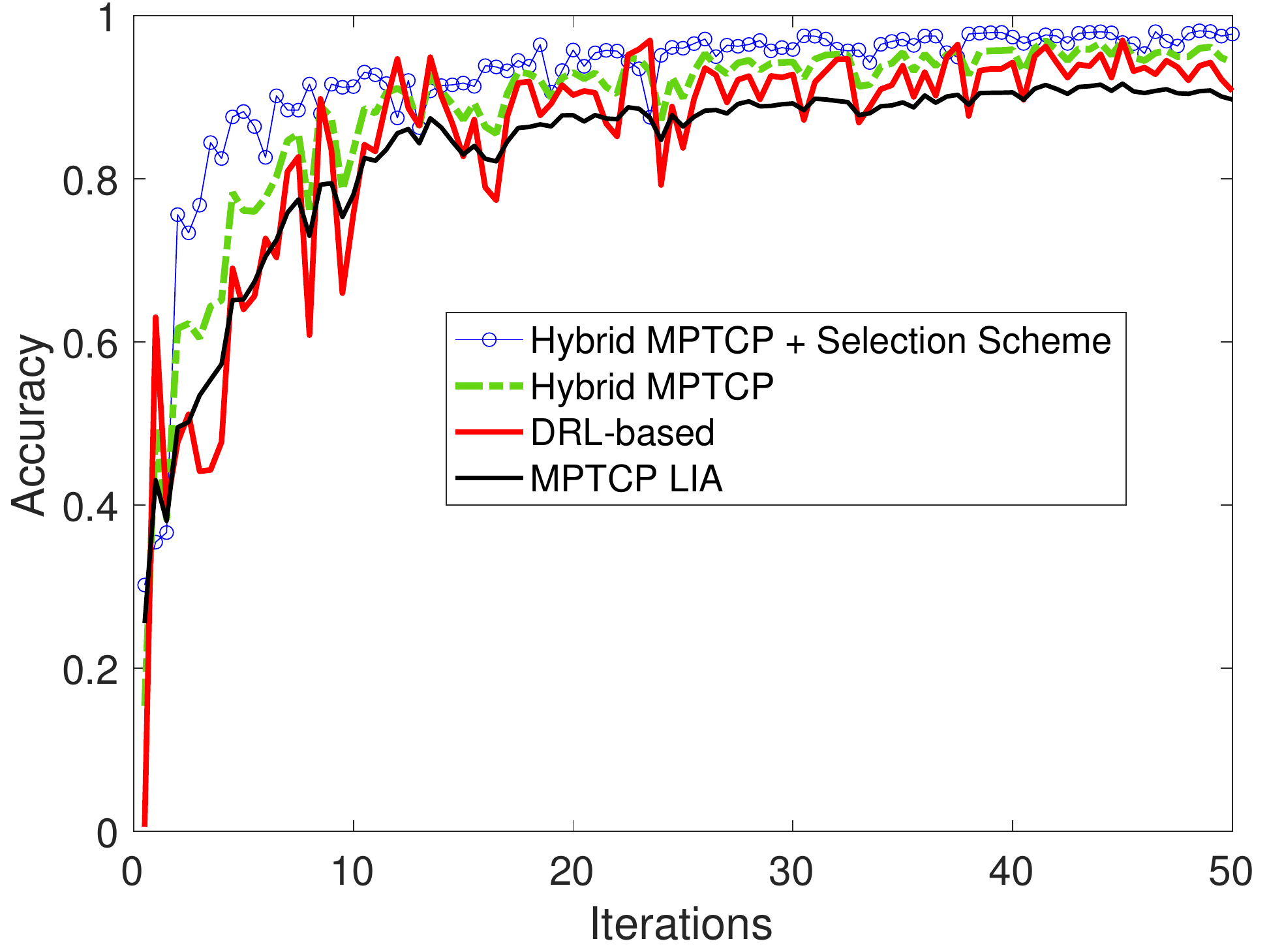}
    \caption{\rm {\color{black}\textbf{Experiment 7}: Comparative DEL accuracy (test accuracy of local model updates) and convergence of different MPTCP with hybrid MPTCP (with/without client selection scheme). With repeated experiments, we observe the average accuracy of all three different versions of MPTCP (DRL-based, MPTCP LIA,
and Hybrid MPTCP).}}
    \label{fig:my_labellast}
    \vspace{-4 mm}
\end{figure}

\subsubsection{Experiment 6 \& 7} {\color{black} In the new Experiment 6, we aim to quantify the impact of increase in the number of stragglers on the DEL iteration time. It is shown below in Fig.~\ref{fig:fig102s} that the average iteration time of DEL increases exponentially with the increase in the number of stragglers. We can see that the proposed Hybrid MPTCP outperforms the state-of-the-art MPTCP protocols in terms of ameliorating the impact of increase in number of stragglers. 

In Experiment 7, to quantify the accuracy (test accuracy of local updates) and the convergence of the DEL process with the proposed DRL agent and compare it across the three MPTCP variants (DRL-based, MPTCP LIA, Hybrid MPTCP), we maintain the iterations logs of the same settings (with a straggler) and observe it for 50 iterations (and average it for all fifty experimental runs). As shown in Fig.~\ref{fig:my_labellast}, we found that the proposed hybrid MPTCP demonstrates higher accuracy and faster convergence compared to the model-based and DRL-based schemes .} 

\subsubsection{Experiment 8}
To understand the overhead of the learning process of DRL agent and compare it across the three MPTCP variants (DRL-based, MPTCP LIA, Hybrid MPTCP), we maintain the CPU utilization (capacity is boosted) logs of the same settings of Experiment 4 and observe it for 150 seconds (to reduce the impact of start-up phases required by the MPTCP algorithms we exclude first 10 seconds of the experiment for all three experimental runs). As shown in Fig.~\ref{fig:fig8}, our hybrid DRL agent incurs a very low overhead compared to the DRL-based schemes. The CPU utilization of Hybrid MPTCP is approximately four times lower than that of the DRL-based. We anticipate that the final optimized kernel version of Hybrid MPTCP could potentially achieve a much lower overhead.

\subsection{Benefits of the Hybrid MPTCP Scheme}
\subsubsection{Efficient and Accelerated Training} 
The inner loop consisting of the classic model-based MPTCP control mechanism significantly contracts the action space of the DRL agent in the case of the proposed hybrid MPTCP. Therefore, this two layering approach has efficiently stimulated the learning of a more general model and much faster in the training phase than the usual DRL-based MPTCP. We compare the learning score of the hybrid MPTCP with the existing DRL-based latter in the performance evaluation section.

\subsubsection{Reduced Computation Overhead} 
Compared to the full learning-based DRL MPTCP, another main advantage of the hybrid MPTCP is the improved CPU utilization and computational complexity. 

\subsubsection{Pros of both Model-based and Learning-based design} Traditional MPTCPs are formulaic and predictable but are not adaptive to various network conditions and suffers from performance impairments. Learning-based MPTCP, by sharp contrast, can learn varying network conditions easily even though they are not easily predictable. The two-level control obtained by leveraging the best model-based and learning-based approach can improve performance in varying network conditions.

{\color{black}
\subsubsection{Improved Convergence with Probing}
The proposed Hybrid MPTCP deliberately adds disturbances to the DRL agent`s outcomes by using model-based MPTCP's continuous adjustments to the windows. These disturbances work as a backup probing  for minimizing the probability of settling down to a wrong equilibrium. Indeed, observe the marginal gain of Hybrid MPTCP in terms of iteration time and fairness (in Figs. 11, 12 and 13) when compared to others in Fig.~\ref{fig:short2}.

 \begin{figure}[t]
\centering 
\includegraphics[width=2.0367825 in]{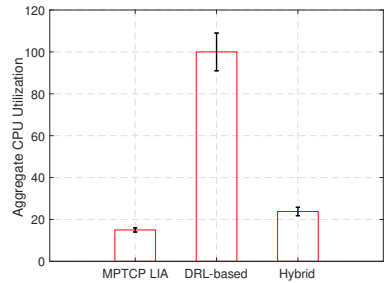}
\caption{\rm \textbf{Experiment 8}: DEL data flows with multipath TCP and single-path TCP contending with each other. We maintain the CPU utilization  logs of the same settings of Experiment 4 and observe them for 150 seconds. We  use five data flows over MPTCP, which are sharing a bottleneck path with rival five data flows over TCP. With repeated experiments, we observe the average CPU utilization in \% of all three different versions of MPTCP (DRL-based, MPTCP LIA,
and Hybrid MPTCP). We can see that with the DRL-based, more than 100\% CPU utilization in a multi-core system demonstrates that it utilizes the second core. \label{fig:fig8}} 
\vspace{-4 mm}
\end{figure}

\subsubsection{Synergy with Client Selection Mechanism} We have assumed heterogeneous workers with different computation and communication capabilities and our main focus in this work is to resolve the challenges  from the lens of transport layer. Moving ahead, we have evaluated the synergy of coupling the proposed hybrid MPTCP with the client selection mechanism. In Experiment~7, by combining the Hybrid MPTCP with an \textit{ideal client selection scheme} (a simple approach which select all pioneers at each
round for the local updates and these updates are aggregated to form a global update by adopting the idea similar to~\cite{li2020federated}). As shown in Fig.~\ref{fig:my_labellast}, we find that the accuracy and convergence improves significantly with the client selection mechanism. This observation requires further investigations  and design from a cross-layer prospective, which will be investigated in the future.
\subsection{Discussion of relevant works}
High-performance MPTCP design is more complex than it appears at first glance. A celebrated single-path TCP design, Performance Oriented Congestion Control (PCC) and its enhancements~\cite{dong2018pcc}, were appealing because it relaxes assumptions about the network and has demonstrated good application-agnostic performance. To begin, one may think of merely executing such application-oriented cutting-edge TCP (e.g., PCC~\cite{dong2018pcc}, BBR~\cite{cardwell2019bbrv2}) on each MPTCP subflows, however, it creates unfairness across the competing TCP and MPTCP flows.  Indeed, Arun \textit{et al.}~\cite{arun2022starvation} have demonstrated that almost all of the application-agnostic delay-bounding designs (designed to ensure good inter-flow fairness) geared at high performance fail to prevent starvation (an extreme case of unfairness).  Additionally, because MPTCP's existing designs~\cite{raiciu2011coupled, peng2016multipath} are inextricably linked to its model-based approach, extending MPTCP's architecture for improvements with the modern TCP theory is notoriously difficult. To this end, MPCC~\cite{gilad2020mpcc}, a multipath extension of PCC-Vivace~\cite{dong2018pcc}, and a few others~\cite{9444785} have demonstrated the underlying MPTCP design challenges considerably. 

With relevant insights from~\cite{gilad2020mpcc, 9444785, arun2022starvation}, we find that none of the existing MPTCP algorithms are capable of generalizing to DEL needs. Indeed, most of them struggles to i) ameliorate unfairness~\cite{arun2022starvation} and ii) generalize well across varying network settings. Rather than offering yet another new MPTCP scheme, we provided a fresh design perspective and illustrated how DRL can learn to greatly improve the performance of the classic model-based MPTCP designs.}
\section{Conclusion}
In this paper, we studied the impending problem of fairness and efficiency in data parallelism for distributed edge learning. We developed a DRL agent coupled tightly with a multipath TCP based bandwidth allocation algorithm, namely Hybrid MPTCP, that interacts with the application level constraints to ameliorate the data parallelism issues (e.g., straggler problem, convergence issues, improve efficiency, unfairness etc.).   The Hybrid MPTCP design consists of two main parts, the DRL agent for adaptive learning and a model-based approach to enhance the design's responsiveness. The developed architecture of the Hybrid MPTCP performs i) interactions with the application and network layers, ii) learning for efficiently allocating bandwidths to all DEL data flows and subflows, ii) harnessing paths bandwidths efficiently by considering the application and flow-level fairness efficaciously.  We have extensively compared the performance of Hybrid MPTCP with the existing model-based and DRL-based data transport algorithms under the same DEL system settings. 

\bibliographystyle{IEEEtran}
\bibliography{shiva}

\begin{thebibliography}{10}
\providecommand{\url}[1]{#1}
\csname url@samestyle\endcsname
\providecommand{\newblock}{\relax}
\providecommand{\bibinfo}[2]{#2}
\providecommand{\BIBentrySTDinterwordspacing}{\spaceskip=0pt\relax}
\providecommand{\BIBentryALTinterwordstretchfactor}{4}
\providecommand{\BIBentryALTinterwordspacing}{\spaceskip=\fontdimen2\font plus
\BIBentryALTinterwordstretchfactor\fontdimen3\font minus
  \fontdimen4\font\relax}
\providecommand{\BIBforeignlanguage}[2]{{%
\expandafter\ifx\csname l@#1\endcsname\relax
\typeout{** WARNING: IEEEtran.bst: No hyphenation pattern has been}%
\typeout{** loaded for the language `#1'. Using the pattern for}%
\typeout{** the default language instead.}%
\else
\language=\csname l@#1\endcsname
\fi
#2}}
\providecommand{\BIBdecl}{\relax}
\BIBdecl

\bibitem{luo2018parameter}
L.~Luo, J.~Nelson, L.~Ceze, A.~Phanishayee, and A.~Krishnamurthy, ``Parameter
  hub: a rack-scale parameter server for distributed deep neural network
  training,'' in \emph{Proceedings of the ACM Symposium on Cloud Computing},
  2018, pp. 41--54.

\bibitem{park2021communication}
J.~Park \emph{et~al.}, ``Communication-efficient and distributed learning over
  wireless networks: Principles and applications,'' \emph{Proceedings of the
  IEEE}, 2021.

\bibitem{9261995}
C.~T. Dinh \emph{et~al.}, ``Federated learning over wireless networks:
  Convergence analysis and resource allocation,'' \emph{IEEE/ACM Transactions
  on Networking}, vol.~29, no.~1, pp. 398--409, 2021.

\bibitem{pokhrel2020federated}
S.~R. Pokhrel and J.~Choi, ``Federated learning with blockchain for autonomous
  vehicles: Analysis and design challenges,'' \emph{IEEE Transactions on
  Communications}, vol.~68, no.~8, pp. 4734--4746, 2020.

\bibitem{park2019distilling}
J.~Park \emph{et~al.}, ``Distilling on-device intelligence at the network
  edge,'' \emph{arXiv preprint arXiv:1908.05895}, 2019.

\bibitem{Verbraeken2020}
\BIBentryALTinterwordspacing
J.~Verbraeken \emph{et~al.}, ``A survey on distributed machine learning,''
  \emph{ACM Comput. Surv.}, vol.~53, no.~2, Mar. 2020. [Online]. Available:
  \url{https://doi.org/10.1145/3377454}
\BIBentrySTDinterwordspacing

\bibitem{8756095}
W.~{Aljoby} \emph{et~al.}, ``On {SDN}-enabled online and dynamic bandwidth
  allocation for stream analytics,'' \emph{IEEE Journal on Selected Areas in
  Communications}, vol.~37, no.~8, pp. 1688--1702, 2019.

\bibitem{9320519}
W.~Li, D.~Liu, K.~Chen, K.~Li, and H.~Qi, ``Hone: Mitigating stragglers in
  distributed stream processing with tuple scheduling,'' \emph{IEEE
  Transactions on Parallel and Distributed Systems}, vol.~32, no.~8, pp.
  2021--2034, 2021.

\bibitem{chilimbi2014project}
T.~Chilimbi \emph{et~al.}, ``Project adam: Building an efficient and scalable
  deep learning training system,'' in \emph{11th $\{$USENIX$\}$ Symposium on
  Operating Systems Design and Implementation ($\{$OSDI$\}$ 14)}, 2014, pp.
  571--582.

\bibitem{bitar2020stochastic}
R.~Bitar, M.~Wootters, and S.~El~Rouayheb, ``Stochastic gradient coding for
  straggler mitigation in distributed learning,'' \emph{IEEE Journal on
  Selected Areas in Information Theory}, vol.~1, no.~1, pp. 277--291, 2020.

\bibitem{yu2002design}
H.~Yu and A.~Vahdat, ``Design and evaluation of a conit-based continuous
  consistency model for replicated services,'' \emph{ACM Transactions on
  Computer Systems (TOCS)}, vol.~20, no.~3, pp. 239--282, 2002.

\bibitem{zhao2019dynamic}
X.~Zhao, A.~An, J.~Liu, and B.~X. Chen, ``Dynamic stale synchronous parallel
  distributed training for deep learning,'' in \emph{2019 IEEE 39th
  International Conference on Distributed Computing Systems (ICDCS)}.\hskip 1em
  plus 0.5em minus 0.4em\relax IEEE, 2019, pp. 1507--1517.

\bibitem{hsieh2017gaia}
K.~Hsieh \emph{et~al.}, ``Gaia: Geo-distributed machine learning approaching
  $\{$LAN$\}$ speeds,'' in \emph{14th $\{$USENIX$\}$ Symposium on Networked
  Systems Design and Implementation ($\{$NSDI$\}$ 17)}, 2017, pp. 629--647.

\bibitem{han2015giraph}
M.~Han and K.~Daudjee, ``Giraph unchained: Barrierless asynchronous parallel
  execution in pregel-like graph processing systems,'' \emph{Proceedings of the
  VLDB Endowment}, vol.~8, no.~9, pp. 950--961, 2015.

\bibitem{tang2020communication}
Z.~Tang \emph{et~al.}, ``Communication-efficient distributed deep learning: A
  comprehensive survey,'' \emph{arXiv preprint arXiv:2003.06307}, 2020.

\bibitem{ouyang2020communication}
S.~Ouyang, D.~Dong, Y.~Xu, and L.~Xiao, ``Communication optimization strategies
  for distributed deep learning: A survey,'' \emph{arXiv preprint
  arXiv:2003.03009}, 2020.

\bibitem{pokhrel2018modeling}
S.~R. Pokhrel and C.~Williamson, ``{Modeling Compound {TCP} Over {WiFi} for
  {IoT}},'' \emph{IEEE/ACM Transactions on Networking (TON)}, vol.~26, no.~2,
  pp. 864--878, 2018.

\bibitem{pokhrel2019fair}
S.~R. Pokhrel \emph{et~al.}, ``{Fair Coexistence of Regular and Multipath {TCP}
  over Wireless Last-Miles},'' \emph{IEEE Transactions on Mobile Computing},
  vol.~18, no.~3, pp. 574--587, 2019.

\bibitem{7864463}
S.~R. Pokhrel, M.~Panda, and H.~L. Vu, ``Analytical modeling of multipath tcp
  over last-mile wireless,'' \emph{IEEE/ACM Transactions on Networking},
  vol.~25, no.~3, pp. 1876--1891, 2017.

\bibitem{xia2019rethinking}
J.~Xia \emph{et~al.}, ``Rethinking transport layer design for distributed
  machine learning,'' in \emph{Proceedings of the 3rd Asia-Pacific Workshop on
  Networking 2019}, 2019, pp. 22--28.

\bibitem{9444785}
S.~R. Pokhrel and A.~Walid, ``Learning to harness bandwidth with multipath
  congestion control and scheduling,'' \emph{IEEE Transactions on Mobile
  Computing}, pp. 1--1, 2021.

\bibitem{pokhrel2020multipath}
S.~R. {Pokhrel} and S.~{Garg}, ``{Multipath Communication With {Deep}
  {Q-Network} For {Industry} 4.0 Automation and Orchestration},'' \emph{IEEE
  Transactions on Industrial Informatics}, pp. 1--1, 2020.

\bibitem{shafigh2006dynamic}
A.~R.~S. Shafigh, F.~Noroozi, and Z.~Khodabandeh, ``Dynamic bandwidth
  allocation with minimum long fluctuations,'' in \emph{2006 8th International
  Conference Advanced Communication Technology}, vol.~1.\hskip 1em plus 0.5em
  minus 0.4em\relax IEEE, 2006, pp. 5--pp.

\bibitem{wen2020joint}
D.~Wen, M.~Bennis, and K.~Huang, ``Joint parameter-and-bandwidth allocation for
  improving the efficiency of partitioned edge learning,'' \emph{IEEE
  Transactions on Wireless Communications}, vol.~19, no.~12, pp. 8272--8286,
  2020.

\bibitem{peng2016multipath}
Q.~Peng, A.~Walid, J.~Hwang, and S.~H. Low, ``{Multipath {{TCP}}: Analysis,
  Design, and Implementation},'' \emph{IEEE/ACM Transactions on Networking},
  pp. 596--609, 2016.

\bibitem{cardwell2019bbrv2}
N.~Cardwell \emph{et~al.}, ``{BBR}v2: A model-based congestion control,'' in
  \emph{Presentation in ICCRG at IETF 104th meeting}, 2019.

\bibitem{raiciu2011coupled}
C.~Raiciu, M.~Handley, and D.~Wischik, ``{Coupled Congestion Control for
  Multipath Transport Protocols, {IETF RFC} 6356},'' 2011.

\bibitem{li2019smartcc}
W.~Li \emph{et~al.}, ``{{SmartCC}: A Reinforcement Learning Approach for
  Multipath {TCP} Congestion Control in Heterogeneous Networks},'' \emph{IEEE
  Journal on Selected Areas in Communications}, 2019.

\bibitem{xu2019experience}
Z.~Xu \emph{et~al.}, ``{Experience-driven Congestion Control: When Multi-Path
  {TCP} Meets Deep Reinforcement Learning},'' \emph{IEEE Journal on Selected
  Areas in Communications}, 2019.

\bibitem{zhang2020machine}
T.~Zhang and S.~Mao, ``Machine learning for end-to-end congestion control,''
  \emph{IEEE Communications Magazine}, vol.~58, no.~6, pp. 52--57, 2020.

\bibitem{mnih2015human}
V.~Mnih \emph{et~al.}, ``{Human-level Control Through Deep Reinforcement
  Learning},'' \emph{Nature}, vol. 518, no. 7540, p. 529, 2015.

\bibitem{lillicrap2015continuous}
T.~P. Lillicrap \emph{et~al.}, ``{Continuous Control With Deep Reinforcement
  Learning},'' Jan.~26 2017, {US} Patent App. 15/217,758.

\bibitem{pokhrel2017analytical}
S.~R. Pokhrel, M.~Panda, and H.~L. Vu, ``Analytical modeling of multipath tcp
  over last-mile wireless,'' \emph{IEEE/ACM Transactions on networking},
  vol.~25, no.~3, pp. 1876--1891, 2017.

\bibitem{mazumdar1991fairness}
R.~Mazumdar, L.~G. Mason, and C.~Douligeris, ``{Fairness in Network Optimal
  Flow Control: Optimality of Product Forms},'' \emph{IEEE Transactions on
  communications}, vol.~39, no.~5, pp. 775--782, 1991.

\bibitem{kelly1998rate}
F.~P. Kelly, A.~K. Maulloo, and D.~K. Tan, ``Rate control for communication
  networks: shadow prices, proportional fairness and stability,'' \emph{Journal
  of the Operational Research society}, vol.~49, no.~3, pp. 237--252, 1998.

\bibitem{winstein2013tcp}
K.~Winstein and H.~Balakrishnan, ``{{TCP} ex machina: Computer-generated
  Congestion Control},'' \emph{ACM SIGCOMM`13}, pp. 123--134, 2013.

\bibitem{hester2018deep}
T.~Hester \emph{et~al.}, ``{Deep Q-learning From Demonstrations},'' in
  \emph{Thirty-Second AAAI Conference on Artificial Intelligence}, 2018.

\bibitem{netravali2015mahimahi}
R.~Netravali, A.~Sivaraman, S.~Das, A.~Goyal, K.~Winstein, J.~Mickens, and
  H.~Balakrishnan, ``Mahimahi: Accurate record-and-replay for $\{$HTTP$\}$,''
  in \emph{2015 {USENIX} Annual Technical Conference ({USENIX}{ATC} 15)}, 2015,
  pp. 417--429.

\bibitem{li2020federated}
T.~Li, A.~K. Sahu, M.~Zaheer, M.~Sanjabi, A.~Talwalkar, and V.~Smith,
  ``Federated optimization in heterogeneous networks,'' \emph{Proceedings of
  Machine Learning and Systems}, vol.~2, pp. 429--450, 2020.

\bibitem{dong2018pcc}
M.~Dong, T.~Meng, D.~Zarchy, E.~Arslan, Y.~Gilad, B.~Godfrey, and M.~Schapira,
  ``{PCC} vivace:{Online-Learning} congestion control,'' in \emph{15th USENIX
  Symposium on Networked Systems Design and Implementation (NSDI 18)}, 2018,
  pp. 343--356.

\bibitem{arun2022starvation}
V.~Arun, M.~Alizadeh, and H.~Balakrishnan, ``Starvation in end-to-end
  congestion control,'' in \emph{Proceedings of the ACM SIGCOMM 2022
  Conference}, 2022, pp. 177--192.

\bibitem{gilad2020mpcc}
T.~Gilad, N.~Rozen-Schiff, P.~B. Godfrey, C.~Raiciu, and M.~Schapira, ``Mpcc:
  online learning multipath transport,'' in \emph{Proceedings of the 16th
  International Conference on emerging Networking EXperiments and
  Technologies}, 2020, pp. 121--135.

\end{thebibliography}
\begin{IEEEbiography}[{\includegraphics[width=1in, height=1.25in]{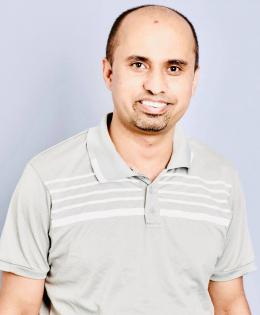}}]
 {Shiva Raj Pokhrel}(SM' 21) was born in Damak, Nepal. He is a Lecturer of Mobile Computing at Deakin University, Geelong, Australia. He received the B.E./M.E. degree  in 2007/2013 from Pokhara University, Nepal and a PhD degree in 2017 from Swinburne University of Technology, Australia.  He was a research fellow at the University of Melbourne and a network engineer at Nepal Telecom (2007-2014). His research interests include multi-connectivity, federated learning, industry 4.0 automation, blockchain modelling, optimization, recommender systems, 6G, cloud computing, dynamics control, Internet of Things and cyber-physical systems as well as their applications in smart manufacturing, autonomous vehicles and  cities. He serves/served as the Workshop Chair/Publicity Co-Chair for several IEEE/ACM conferences including IEEE INFOCOM, IEEE GLOBECOM, IEEE ICC, ACM MobiCom, and more.  He was a recipient of the prestigious  Marie Curie Fellowship in 2017-2020 and the finalist of the IEEE Future Networks' \textit{Connecting The Unconnected Challenge} (CTUC) in 2021.
 \end{IEEEbiography}

\begin{IEEEbiography}[{\includegraphics[width=1in,height=1.25in,keepaspectratio]{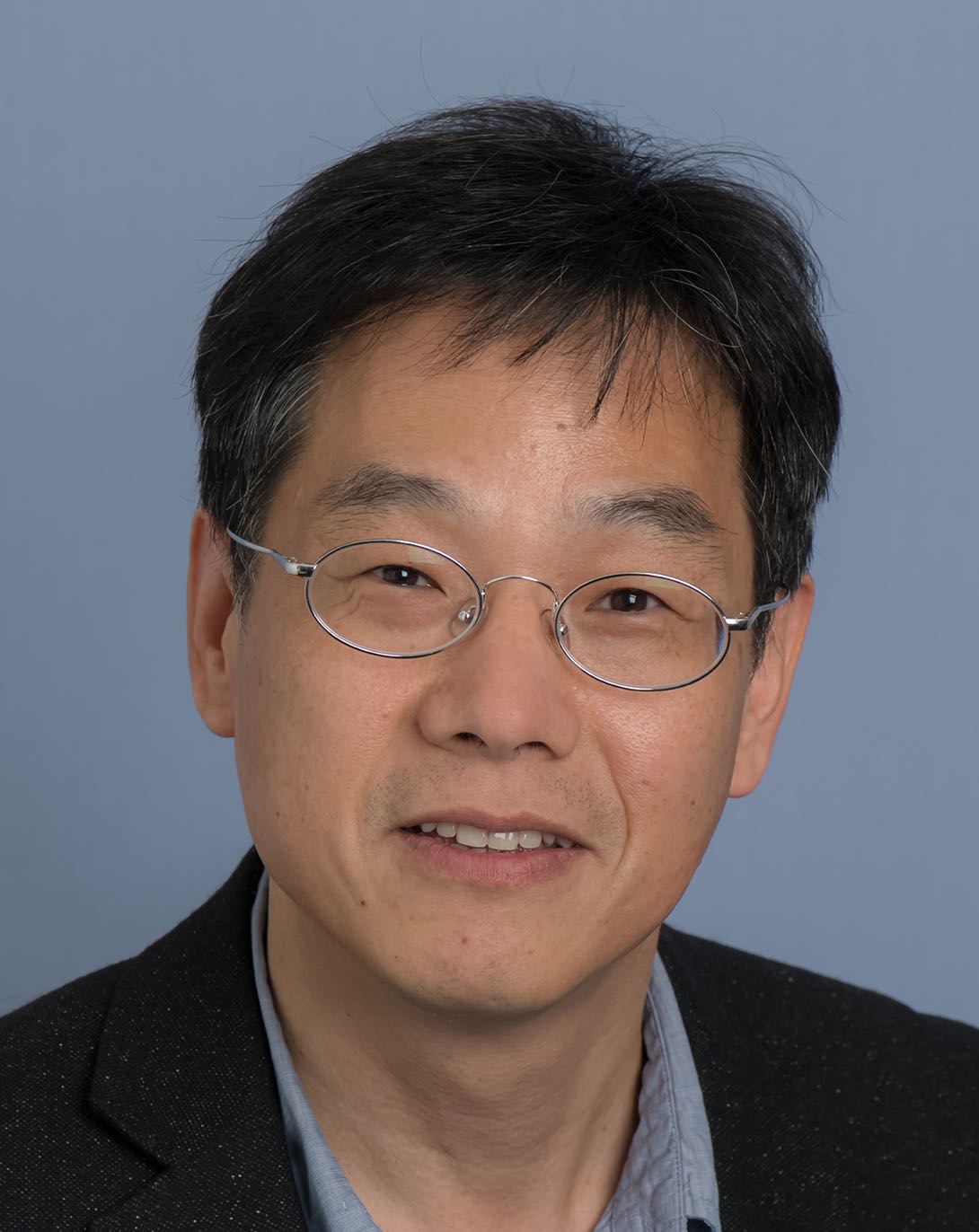}}]
{Jinho~Choi}
(SM' 02) was born in Seoul, Korea. He received B.E. (magna cum laude) degree in electronics engineering in 1989 from Sogang University, Seoul, and M.S.E. and Ph.D. degrees in electrical engineering from Korea Advanced Institute of Science and Technology (KAIST) in 1991 and 1994, respectively. He is with the School of Information Technology, Burwood, Deakin University, Australia, as a Professor. Prior to joining Deakin in 2018, he was with Swansea University, United Kingdom, as a Professor/Chair in Wireless, and Gwangju Institute of Science and Technology (GIST), Korea, as a Professor. His research interests include the Internet of Things (IoT), wireless communications, and statistical signal processing. He authored two books published by Cambridge University Press in 2006 and 2010. Prof. Choi received a number of best paper awards including the 1999 Best Paper Award for Signal Processing from EURASIP. He is on the list of World’s Top 2\% Scientists by Stanford University in 2020 and 2021. Currently, he is an Editor of IEEE Wireless Communications Letters and a Division Editor of Journal of Communications and Networks (JCN). He has also served as an Associate Editor or Editor of other journals including IEEE Trans. Communications, IEEE Communications Letters, JCN, IEEE Trans. Vehicular Technology, and ETRI journal.

\end{IEEEbiography}

\begin{IEEEbiography}[{\includegraphics[width=1in, height=1.25in]{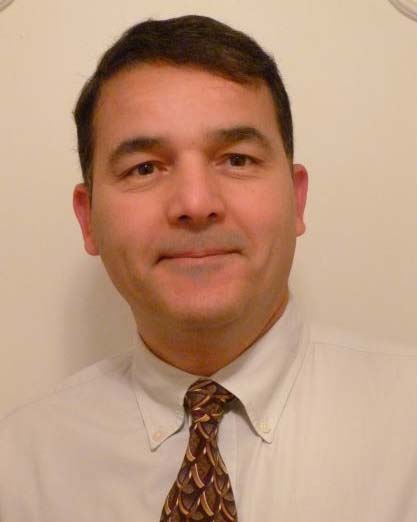}}]
{Anwar Walid}(S'89-M'90-SM'02-F'09) received the B.S. and M.S. degrees in
electrical and computer engineering from New York
University, New York City, NY, USA, and the Ph.D.
degree from Columbia University, New York City,
NY, USA. He was at Nokia Bell Labs, Murray
Hill, NJ, USA, as the Head of the Mathematics of
System Research Department and as the Director
of University Research Partnerships. He is currently
the Director of Network Intelligence and Distributed
Systems Research and a Distinguished Member of
the Research Staff at Nokia Bell Labs. He is also
an Adjunct Professor at the Electrical Engineering Department, Columbia
University. He has over 20 U.S. and international granted patents on various
aspects of networking and computing. His research interests are in the control
and optimization of distributed systems, learning models and algorithms with
applications to Internet of Things (IoT), digital health, smart transportation,
cloud computing, and software-defined networking. He is a fellow of the
IEEE, and an Elected Member of the International Federation for Information
Processing Working Group 7.3 and the Tau Beta Pi Engineering
Honor Society. He received awards from the IEEE and ACM, including
the 2017 IEEE Communications Society William R. Bennett Prize and
the ACM SIGMETRICS/IFIP Performance Best Paper Award. He served
as an Associate Editor for the IEEE/ACM TRANSACTIONS ON CLOUD
COMPUTING, IEEE Network Magazine, and the IEEE/ACM TRANSACTIONS
ON NETWORKING. He served as the Technical Program Chair for the IEEE
INFOCOM, as the General Chair for the 2018 IEEE/ACM Conference on
Connected Health (CHASE), and as a Guest Editor for the IEEE IoT Journal—
Special Issue on AI-Enabled Cognitive Communications and Networking
for IoT.
\end{IEEEbiography}
 \end{document}